\documentclass[%
 reprint,
 amsmath,amssymb,
 aps,nofootinbib
]{revtex4-1}
\usepackage[utf8]{inputenc}
\usepackage{xcolor}
\usepackage{graphicx}
\usepackage{dcolumn}
\usepackage{bm}
\usepackage{mathtools}
\usepackage{xcolor}
\usepackage{natbib}
\usepackage{graphicx}
\usepackage{amsmath}

\newcommand{\equ}[1]{\begin{eqnarray}#1\end{eqnarray}}
\newcommand{\wide}[1]{\begin{widetext}#1\end{widetext}}
\newcommand{\tianshu}[1]{\textcolor{blue}{#1}}

\newcommand{\apjl}{Astrophysical Journal, Letters}
\newcommand{\mnras}{Monthly Notices of the RAS}
\newcommand{\apjs}{Astrophysical Journal, Supplement}
\newcommand{\aap}{Astronomy and Astrophysics}
\newcommand{\ssr}{Space Science Reviews}

\usepackage{comment}

\begin{document}
\title[Generalized]{A Generalized Kompaneets Formalism for Inelastic Neutrino-Nucleon Scattering in Supernova Simulations}

\author{Tianshu Wang}
\email{tianshuw@princeton.edu}
\affiliation{Department of Astrophysical Sciences, Princeton University, Princeton, NJ 08544}
\author{Adam Burrows}
\affiliation{Department of Astrophysical Sciences, Princeton University, Princeton, NJ 08544}

\begin{abstract}
Based on the Kompaneets approximation, we develop a robust methodology 
to calculate spectral redistribution via inelastic neutrino-nucleon 
scattering in the context of core-collapse supernova simulations.  The resulting equations 
conserve lepton number to machine precision and scale linearly, not quadratically, 
with number of energy groups. The formalism also provides an elegant means to
derive the rate of energy transfer to matter which, as it must, automatically 
goes to zero when the neutrino radiation field is in thermal equilibrium. Furthermore, we derive the next-higher-order in $\varepsilon/mc^2$ correction to the neutrino Kompaneets equation.
Unlike other Kompaneets schema, ours also generalizes to the case of
anisotropic angular distributions, while retaining the conservative form
that is a hallmark of the classical Kompaneets equation. Our formalism enables 
immediate incorporation into supernova codes that follow the spectral angular 
moments of the neutrino radiation fields.
\end{abstract}


\maketitle
\section{Introduction}
\label{intro}


Due to substantial progress over the last decade on many fronts, the 
general viability of the neutrino heating mechanism of core-collapse 
supernovae (CCSNe) \citep{1985ApJ...295...14B,Bethe:1990mw} has been put on a firmer foundation.  Though most 
spherical (1D) models do not explode, most, though not all, two-dimensional 
(2D) axisymmetric and three-dimensional (3D) models incorporating sophisticated 
neutrino physics, state-of-the-art numerical tools, realistic 
nuclear equations of state (EOSes), and detailed massive-star progenitor models do explode 
\citep{melson:15a,lentz:15,TaKoSu16,roberts:16,2016ApJ...818..123B,MuMeHe17,OcCo18,oconnor_couch2018,ott2018_rel,burrows_low_2019,vartanyan2019,2019ApJ...873...45G,2020MNRAS.491.2715B}.  Despite the resource intensity of many of
these complex models, theorists are now able to explore suites of 
multi-dimensional simulations and map the parameter dependencies
of the remaining ambiguities. What has emerged is a nuanced understanding
of the factors of explosion.  The latter include sufficient neutrino heating behind
a temporarily-stalled bounce shock wave as the direct agency of explosive power;
the crucial role of anisotropic turbulence in augmenting the driving stress 
behind the shock \citep{bhf1995}; the breaking of spherical symmetry to allow simultaneous
accretion and explosion (the former to ensure the continuance of 
sufficient driving neutrino emissions from the residual proto-neutron star (PNS),
despite the reversal of infall implied by explosion); and core progenitor 
structures that are conducive to eventual explosive instability.  In addition,
the potential roles of many-body neutrino-matter interactions 
\citep{1995PhRvD..51.6635K,1998PhRvC..58..554B,sawyer1999,roberts2012,roberts_reddy2017,horowitz:17,burrows:18}, of turbulence in the
progenitor cores themselves \citep{muller_janka_pert,CoChAr15,jones_2016,2016ApJ...822...61C,muller2017}, 
of remaining uncertainties in the nuclear EOS \citep{steiner:13,schneider_eos_2019}, and   
of rotation \citep{TaKoSu16,summa2018} and magnetic fields \citep{burrows2007_mag,MoRiOt14,2020arXiv200302004K} in a subset of massive-star explosions
continue to exercise the community. At least as important, the precise mapping between 
progenitor structure and outcome, importantly including explosion energy, residual
mass (and whether a neutron star or black hole is birthed), recoil kicks, pulsar and 
magnetar magnetic fields, and nucleosynthetic yields, has yet to be convincingly 
determined. Hence, despite the palpable progress claimed above, much remains to 
be done. 

Neutrino heating of material behind a stalled shock itself drives the turbulence \citep{burrows2012}
\footnote{though the so-called ``SASI'' (Standing Accretion Shock Instability)  \citep{blondin2003,foglizzo:07}
can play a subdominant role}, and they together seem central to reversing an accretion shock 
into explosion \footnote{It is in part turbulence and its chaotic character that mitigates against
a simple correspondence between progenitor structure and outcome and makes theoretical 
prediction complex. As a result, current thinking is that Nature provides distribution functions of final state properties 
[explosion energy, kick speed, morphology, nucleosynthesis, proto-neutron star mass, etc.] for a given
progenitor. What these distribution functions may be is a topic of future research. }. Therefore, 
the energy deposition rate due to neutrino-matter interactions in the semi-transparent ``gain region" 
\citep{1985ApJ...295...14B} between the PNS left behind and the shock assumes a pivotal role. The dominant
processes are super-allowed charged-current absorption of electron neutrinos ($\nu_e$)
and anti-electron neutrinos $\bar{\nu}_e$ on nucleons, via the reactions $\nu_e + n \rightarrow e^- + p$
and $\bar{\nu}_e + p \rightarrow  e^+ + n$, respectively \citep{burrows:06}.  Though the cross sections 
for these processes involve some subtleties, they are well understood. However, inelastic 
scattering of neutrinos of all neutrino species on electrons and on nucleons 
\citep{1985ApJS...58..771B,thomp_bur_horvath,rampp_janka2002,2002astro.ph.11404B,burrows_thompson2004,burrows:06} also
heat the matter, though at a lower rate.  However, when a model is near explosion,
even 10\% $-$ 20\% effects can loom large and such is the case here.  The low cross sections
of neutrino-electron scattering coupled with the high average energy transfer to the 
electrons (due to the low electron mass) competes with the high cross section of 
neutrino-nucleon scattering coupled with its correspondingly low recoil energies 
(due to the high nucleon mass). The net result is comparable matter heating rates.
Therefore, it is important to handle inelasticity for both reaction classes in sophisticated transport 
schemes. For neutrino-electron scattering, this involves $N(N-1)/2$ coupled pairs of energy groups,
where $N$ is the number of energy groups used in a calculation, with the result that
a large number of groups becomes quite expensive.  It is this strong scaling with $N$ 
that has limited state-of-the-art 3D simulations to $N$s of, for instance, twelve, with the
result that accuracy may be compromised.

The large energy transfer of neutrino-electron scattering necessitates a large coupling 
matrix.  However, for inelastic neutrino-nucleon scattering the energy transfers are predominantly
small, and smaller than the energy bin widths in viable simulations.  However,
to date this inelasticity has frequently been handled in the same fashion as neutrino-electron
scattering, with the same quadratic penalty. Furthermore, attempts have been made to employ
sub-energy-grid methods to handle the small energy transfers \citep{burrows_thompson2004,burrows:06,chimera}, 
but these are frequently too approximate or don't conserve lepton number by construction.  The result is some ambiguity
in the contribution of neutrino-nucleon scattering inelasticity to the total heating rates.
In addition, the numerical difficulties of including inelasticity in $\nu_{\mu}$, $\nu_{\tau}$,
$\bar{\nu}_{\mu}$ and $\bar{\nu}_{\tau}$ transport, for which its effect 
on the source term can be larger than that due to absorption processes, has resulted in
the dropping of this effect altogether in many otherwise sophisticated supernova codes. 
This is unfortunate, since there is every indication that heating due to neutrino-nucleon
scattering execeeds that due to neutrino-electron scattering \citep{burrows:18,vartanyan2018a}.   

Given this, we have sought to develop a more computationally robust and accurate method
with which to incorporate inelastic scattering off nucleons into neutrino transport
algorithms.  We do this by building on the earlier work of Suwa et al \cite{Suwa2019}. 
Since the energy transfer per scattering is small, neutrino-nucleon
inelasticity can naturally be handled using the Kompaneets small-energy transfer 
ansatz for which the change in the Boltzmann distribution function ($f$) is expanded to 
quadratic order in energy transfer \citep{kom1957}.  This classically is done for photon energy redistribution by Thomson/Compton scattering \citep{sunyaev1972,rybicki_1979}.  Normally, the resulting equation for the rate of change of $f$
assumes that $f$ is isotropic.  We here drop this requirement to derive the generalized Kompaneets
equation for the evolution of $f$ in energy space when $f$ is anisotropic in angle and focus on
the evolution of its angular moments.  Without loss of generality, we ignore the spatial advection
operators, assuming they are operator split off, and focus on the fluxes in neutrino energy space.
The resulting equations elegantly conserve lepton number and provide a direct means to calculate
the heat transfer to matter that naturally goes to zero when $f$ is in thermal equilibrium.
The equations are also linear in the number of energy groups.

\section{Derivation}

In the derivation of our formalism, we use the natural units $\hbar=c=1$, and dimension these constants only when we need physical results. In the co-moving frame, the Boltzmann equation is:
\equ{
\frac{df}{dt}&=&G^2\int \frac{d^3p'}{(2\pi)^3}S(q,\omega)[(1+\mu)V^2+(3-\mu)A^2]\nonumber\\
& &\times\{(1-f)f'e^{-\beta\omega}-f(1-f')\}
\label{equ:def}
}
where $G^2=1.55\times10^{-33}$ cm$^3$ MeV$^{-2}$ s$^{-1}$. $V$ and $A$ are the vector and axial-vector coupling constants. For protons, $V_p=\frac{1}{2}-2\sin^2(\theta_W)$ and $A_p=\frac{1}{2}g_A$, while for neutrons $V_n=-\frac{1}{2}$ and $A_n=-\frac{g_A}{2}$. Here, $\theta_W$ is the Weinberg angle and $\sin^{2}(\theta_W)=0.23129$ and $g_A=-1.2723$ is the axial-vector coupling constant \cite{thomp_bur_horvath,burrows_thompson2004,Suwa2019,PDG_2016}. Note that the second line of Eq. (\ref{equ:def}) is still inside the integrand. In this paper, everything after the integral sign is inside the integrand, no matter if the equation is multi-line or not.

\begin{figure}
    \centering
    \includegraphics[width=0.48\textwidth]{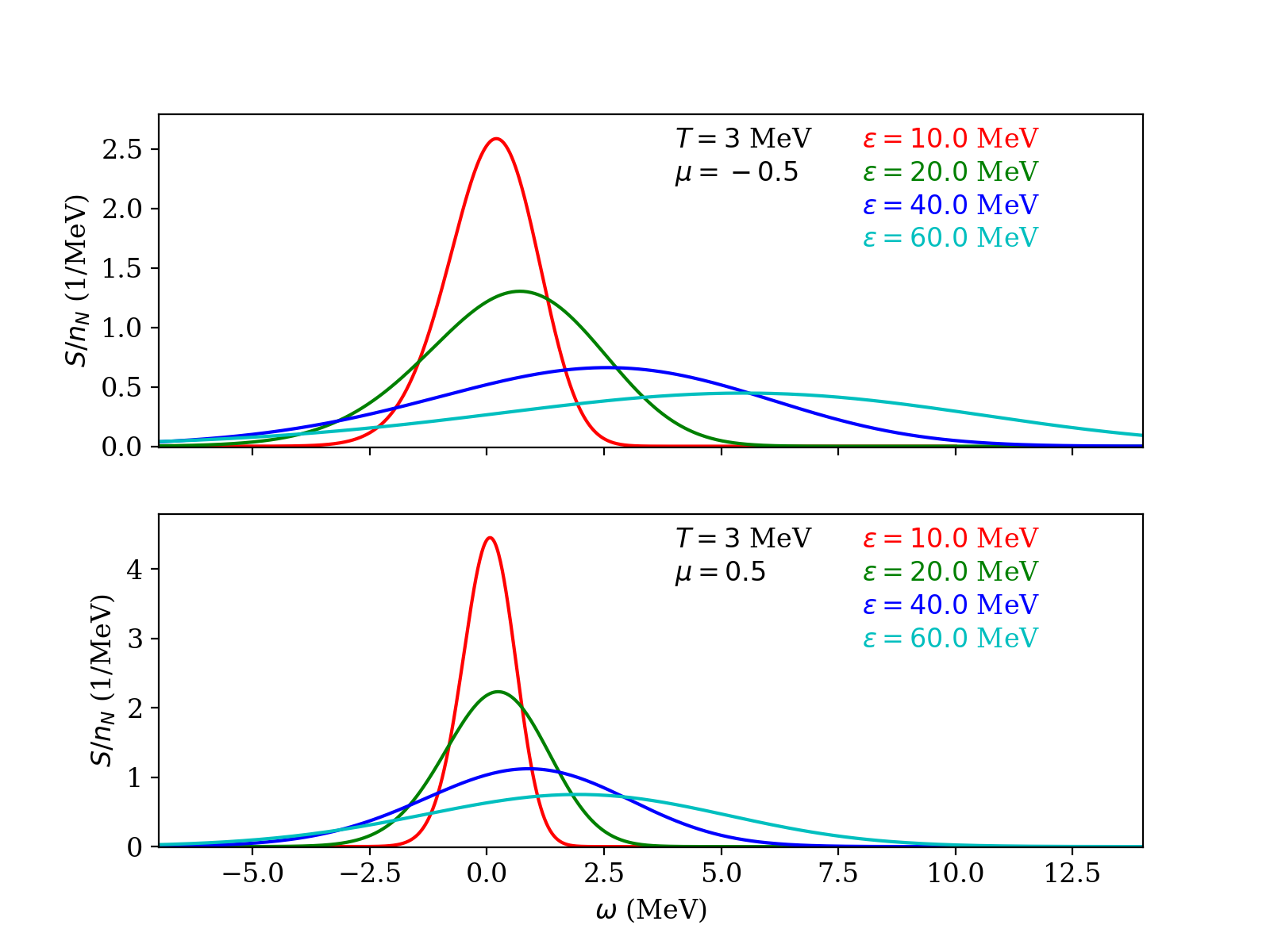}
    \caption{The dynamical structure function versus the neutrino energy transfer for some representative parameters. The temperature here is $T=3$ MeV. Top: $\mu=-0.5$. Bottom: $\mu=0.5$, where $\mu$ is the cosine of the angle between the incoming and outgoing neutrinos. A mass density of 10$^{10}$ g cm$^{-3}$ is assumed. Each line is actually a superposition of the exact (Eq. (\ref{equ:exact})) and approximate (Eq. (\ref{equ:approx})) structure functions.}
    \label{fig:sf}
\end{figure}

In Eq. (\ref{equ:def}), $f=f(\theta,\varepsilon)$ and $f'=f(\theta',\varepsilon')$ are the distribution functions of neutrinos before and after scattering, $\varepsilon$ is the neutrino energy,  $\mu=cos(\theta)cos(\theta')+sin(\theta)sin(\theta')cos(\phi-\phi')$ is the cosine between the incoming and outgoing neutrino momentum vectors, $\omega=\varepsilon-\varepsilon'$ is the energy transfer, $q=\sqrt{\varepsilon^2+\varepsilon'^2-2\varepsilon\varepsilon'\mu}$ is the momentum transfer, $\beta=\frac{1}{kT}$ is the inverse of the nucleon temperature, and $S(q,\omega)$ is the dynamical structure function:
\equ{
S(q,\omega)&=&\frac{1}{1-e^{-\beta\omega}}\frac{m^2}{\pi q\beta}\ln(\frac{1+e^{-Q^2+\eta}}{1+e^{-Q^2+\eta-\beta\omega}})\label{equ:exact}\\
Q&=&\sqrt{\frac{\beta m}{2}}(-\frac{\omega}{q}+\frac{q}{2m})\, ,
}
where $\mu_\nu$ is the nucleon chemical potential and $\eta=\beta\mu_\nu$. The nucleon mass $m$ is $938.3$ MeV for the proton and $939.6$ MeV for the neutron. Here, for convenience we use $m=939$ MeV. In the non-degenerate nucleon limit, the structure function can be approximated by \citep{burrows_thompson2004}:
\equ{
S(q,\omega)
&\approx&\frac{n_N(2\pi m\beta)^{\frac{1}{2}}}{q}e^{-Q^2}\, ,
\label{equ:approx}
}
where $n_N=2(\frac{m}{2\pi \beta})^{3/2}\exp(\eta)$ 
is the nucleon number density. This approximation works very well in the relevant parameter regimes. Figure \ref{fig:sf} renders the shape of structure function with different parameters. From this figure, we see that Eq. (\ref{equ:approx}) is a very good approximation to the exact non-interacting structure function.

We introduce dimensionless variables $x=\beta\varepsilon$ and $x'=\beta\varepsilon'$ and $\alpha=\frac{x-x'}{x}$ and define three new functions to simplify the equation:
\wide{
\equ{
u(\mu)&=&(1+\mu)V^2+(3-\mu)A^2\\
g(x,\theta;x',\theta')&=&(1-f)f'e^{\frac{1}{2}(x'-x)}-f(1-f')e^{\frac{1}{2}(x-x')}\\
\tilde{S}(x,x',\mu)&=&S(q,\omega)e^{-\frac{1}{2}\beta\omega}=\frac{n_N(2\pi\beta m)^{\frac{1}{2}}\beta}{\sqrt{x^2+x'^2-2xx'\mu}}\exp\left\{-\frac{\beta m}{2}\left[\frac{(x-x')^2}{x^2+x'^2-2xx'\mu}+\frac{x^2+x'^2-2xx'\mu}{4\beta^2m^2}\right]\right\}
}
}
We call $\tilde{S}$ the modified structure function.
With these new variables and functions, the Boltzmann equation is
\equ{
\frac{df}{dt}
&=&\frac{G^2x^3}{(2\pi)^3\beta^3}\int d\Omega'\int_{-\infty}^{1}d\alpha \nonumber\\
& &u(\mu) g(x,\theta;x',\theta')\tilde{S}(x,x(1-\alpha),\mu) (1-\alpha)^2
}
In this paper, we regard the neutrino as a massless fermion, so we do not distinguish $p$ and $\varepsilon$.
The total neutrino energy is $E_\nu=\int\frac{d^3p}{(2\pi)^3}pf$, and, thus, the total rate of energy loss to the matter (the energy deposition rate) is 
\equ{
\dot{Q}_\nu&=&-\frac{dE_\nu}{dt}=-\int\frac{d^3p}{(2\pi)^3}p\frac{df}{dt}\, .
}
We can write the right-hand-side of the equation for $\frac{df}{dt}$ using a total derivative:
\equ{
\frac{df}{dt}&=&\frac{1}{x^2}\frac{dI_\nu}{dx}\, .
}
Then, the energy deposition rate is given by
\equ{
\dot{Q}_\nu&=&\frac{1}{(2\pi)^3\beta^3}\int dxd\Omega \dot{q}_\nu\nonumber\\
&=&\frac{1}{(2\pi)^3\beta^4}\int dxd\Omega I_\nu\, ,
}
where $\dot{q}_\nu=\frac{1}{\beta}\left(I_\nu-\frac{d}{dx}(xI_\nu)\right)$ is the spectrum of energy deposition.

We now proceed first to derive the neutrino Kompaneets equation when the neutrino radiation field is isotropic and higher-order corrections in $\varepsilon/mc^2$ are ignored. We then go on in \S\ref{aniso} to address the anisotropic case, and then in \S\ref{higher} derive the higher-energy correction to the neutrino Kompaneets equation in the isotropic case. 

\subsection{Isotropic Case}
We here consider the isotropic case where $f(\varepsilon,\theta)=f(\varepsilon)$ and $g(x,\theta;x',\theta')=g(x;x')$. The angular integral becomes simpler, since $d\Omega'=2\pi d\mu$. 
First, we expand $g(x,x')$ around $g(x,x)$:
\equ{
g(x,x')
&=&\sum_{k=0}^{\infty}\frac{(x'-x)^k}{k!}g^{(k)}(x)\, ,
}
where $g^{(k)}(x)=\frac{\partial^k g(x,x')}{\partial x'^k}\vert_{x'=x}$. The first few $g^{(k)}$ terms are listed in Appendix \ref{app:A}. After substituting into $\frac{df}{dt}$, we obtain
\equ{
\frac{1}{x^2}\frac{dI_\nu}{dx}=\frac{df}{dt}
&=&\frac{G^2x^3}{(2\pi)^2\beta^3}\int_{-1}^1 d\mu 
\nonumber\\
& & u(\mu)\sum_{k=0}^{\infty}g^{(k)}(x)a_{k}(x,\mu)\, ,
\label{equ:Inu_iso}
}
where $a_k(x,\mu)=\int_{-\infty}^1\frac{(-x\alpha)^k}{k!}\tilde{S}(x,x(1-\alpha),\mu)(1-\alpha)^2d\alpha$.

The calculation of $a_k(x,\mu)$ is given in Appendix \ref{sec:ak}. We need only the dominant terms $a_1$ and $a_2$, since $g^{(0)}(x)=0$. The result is Eq. (\ref{equ:ak2}):
\equ{
a_1(x,\mu)&=&\frac{12n_N\pi}{m}\left[1+O\left(\frac{1}{\beta m}\right)\right]\\
a_2(x,\mu)&=&\frac{2xn_N\pi}{m}\left[1+O\left(\frac{1}{\beta m}\right)\right]\, .
}
We then obtain
\equ{
I_\nu&=&\frac{2n_NG^2}{3\pi\beta^3m}(V^2+5A^2)x^6(\frac{df}{dx}+f-f^2)\, .
}

To simplify the expressions derived below we define the following variables \cite{burrows_thompson2004}:
\equ{
\sigma_{\text{tr}}&=&\frac{2G^2}{3\pi\beta^2}(V^2+5A^2)=\frac{2G^2(kT)^2}{3\pi c}(V^2+5A^2)\\
\delta_N&=&\frac{V^2-A^2}{V^2+3A^2}
}
$\sigma_{\text{tr}}$ the momentum transport cross section associated with the transfer of momentum between the scattered neutrino and the nucleon at $\varepsilon = kT$.  Multiplying $\sigma_{\text{tr}}$ by $x^2$ yields the lowest-order expression for the transport cross section at energy $\varepsilon$. $\delta_N$ is the scattering anisotropy factor.

With these substituions, the results are:
\equ{
I_\nu
&=&\frac{kT}{mc^2}\sigma_{\text{tr}} n_Ncx^6(\frac{df}{dx}+f-f^2)\label{equ:iso1}\\
\frac{df}{dt}
&=&\frac{1}{x^2}\frac{dI_\nu}{dx}\nonumber\\
& &=\frac{kT}{mc^2}\sigma_{\text{tr}} n_Nc\frac{1}{x^2}\frac{d}{dx}\left[x^6(\frac{df}{dx}+f-f^2)\right]\label{equ:iso2}\\
\dot{q}_\nu
&=&kT\left(I_\nu-\frac{d}{dx}(I_\nu)\right)\label{equ:iso3}\\
\dot{Q}_\nu
&=&\frac{(kT)^4}{2\pi^2\hbar^3c^3}\int dx I_\nu\, ,
\label{equ:iso4}
}
where $\dot{Q}_\nu$ is the matter heating rate due to inelastic scattering off nucleons. Recall that $x=\beta\varepsilon$.
Eqs. (\ref{equ:iso1})--(\ref{equ:iso4}) encapsulate our neutrino Kompaneets formalism, including a powerful expression for the neutrino-matter energy transfer/heating rate. The total derivative form of this Kompaneets equation ensures that neutrino number is conserved, since total number is proportional to the integral of $x^2\frac{df}{dt}$. $I_\nu$ is proportional to the flux in energy space and the differences between fluxes at energy bin boundaries give the change in a bin particle number exactly. This fact is independent of the size of an energy bin. The above formulae also provide a relation between the rate of energy deposition and the flux $I_\nu$, and ensures, as it should, that when there is no net number flux there is no 
energy transfer to the matter. Importantly, these equations guarantee when the neutrinos are in local thermal equilibrium with the matter at the matter temperature (when $\frac{df}{dx}+f-f^2=0$) that there is not only no number redistribution among the bins, but also that there is no energy transfer to the matter.

\subsection{Anisotropic Case}
\label{aniso}
With minor changes, the above method can also be generalized to the anisotropic neutrino distribution case. Although $\frac{df}{dt}$ can't be written in the form of $\frac{1}{x^2}\frac{dI_\nu}{dt}$, since the particle number is not conserved along any given direction, we can profitably use relations for the angular moments.

Use the same notation as found in \citep{burrows_thompson2004}, we define Legendre expansion terms:
\equ{
{J}_\nu&=&\frac{1}{2}\int_{0}^\pi d\theta' f(\theta',x)\sin(\theta')\\
{H}_\nu&=&\frac{1}{2}\int_{0}^\pi d\theta' f(\theta',x)\sin(\theta')\cos(\theta')\\
{K}_\nu&=&\frac{1}{2}\int_{0}^\pi d\theta' f(\theta',x)\sin(\theta')\times\frac{3}{2}(3\cos^2(\theta')-1)\\
}
and
\equ{
\dot{q}_{\nu,0}&=&\frac{1}{2}\int_{0}^{\pi}d\theta \sin(\theta) \dot{q}_\nu\, .
}

$J_\nu$ is the angle-averaged particle number distribution and satisfies particle number conservation, so it can be written in the following form:
\equ{
\frac{dJ_\nu}{dt}&=&\frac{1}{x^2}\frac{dI_{\nu,0}}{dx}\, .
}
Thus,
\equ{
\dot{q}_{\nu,0}&=&\frac{1}{\beta}\left(I_{\nu,0}-\frac{d}{dx}(xI_{\nu,0})\right)\, .
}
We can calculate $I_{\nu,0}$ in the same way as is done for isotropic case, and then find that $I_{\nu,0}$ is the angle-averaged value of $I_\nu$ given in Eq. (\ref{equ:Inu_iso}):
\equ{
\frac{1}{x^2}\frac{dI_{\nu,0}}{dx}=\frac{dJ_\nu}{dt}
&=&\frac{G^2x^3}{2(2\pi)^3\beta^3}\int \sin(\theta)d\theta d\Omega' 
\nonumber\\
& & u(\mu)\sum_{k=0}^{\infty}g^{(k)}(x)a_{k}(x,\mu)\, .
\label{equ:Inu}
}

Using the expressions summarized in the Appendix \ref{app:A}, we can derive the generalizations:
\wide{
\equ{
I_{\nu,0}
&=&\frac{kT}{mc^2}\sigma_{\text{tr}} n_Ncx^6\left[(\frac{dJ_\nu}{dx}+J_\nu-J_\nu^2)+\frac{3-3\delta_N}{3-\delta_N}H_\nu^2+\frac{2}{27}\frac{3\delta_N}{3-\delta_N}K_\nu^2\right]
\label{equ:kom_aniso}
}
}
We see that the higher-order angular moments ($H_{\nu}$ and $K_{\nu}$) enter quadratically, so their contributions are generically small when the neutrinos are not degenerate ($f\ll 1$).  This will be the case in the important and relevant gain region.

Using our formalism, we can also calculate an expression for $\frac{dH_{\nu}}{dt}$.
First, we start with the expression:
\equ{
\frac{dH_\nu}{dt}&=&\frac{G^2x^3}{2(2\pi)^3\beta^3}\int  \cos(\theta)\sin(\theta)d\theta d\Omega'\nonumber\\
& &u(\mu)\sum_{k=0}^{\infty}g^{(k)}(x)a_{k}(x,\mu)\, .
}
In general, for higher-order angular moments, it is the $g^{0}(x,\theta,\theta')$ term that dominates in $I_{\nu,n}$, since $a_0$ is $\beta m$ times larger than other coefficients. The evolution of $H_\nu$ is then given by:
\equ{
\frac{d{H}_\nu}{dt} 
&=&-\sigma_{\text{tr}} n_Ncx^2{H}_\nu\, .
\label{equ:H}
}
The right-hand-side of this equation is exactly the term expected in the equation that emerges from taking the first-angular moment of the Boltzmann equation.  That equation embodies radiation momentum conservation and momentum transfer to the matter. The spatial derivative term we do not address here (assumed in our formalism to be operator-split off) is the gradient of the pressure tensor and the equality of that term with the right-hand-side of Eq. (\ref{equ:H}) yields the diffusion equation.

In summary, the results are
\wide{
\equ{
I_{\nu,0}
&=&\frac{kT}{mc^2}\sigma_{\text{tr}} n_Ncx^6\left[(\frac{dJ_\nu}{dx}+J_\nu-J_\nu^2)+\frac{3-3\delta_N}{3-\delta_N}H_\nu^2+\frac{2}{27}\frac{3\delta_N}{3-\delta_N}K_\nu^2\right]\\
\frac{dJ_\nu}{dt}
&=&\frac{1}{x^2}\frac{dI_{\nu,0}}{dx}
=\frac{kT}{mc^2}\sigma_{\text{tr}} n_Nc\frac{1}{x^2}\frac{d}{dx}\left\{x^6\left[(\frac{dJ_\nu}{dx}+J_\nu-J_\nu^2)+\frac{3-3\delta_N}{3-\delta_N}H_\nu^2+\frac{2}{27}\frac{3\delta_N}{3-\delta_N}K_\nu^2\right]\right\}\\
\dot{q}_{\nu,0}&=&kT\left(I_{\nu,0}-\frac{d}{dx}(xI_{\nu,0})\right)\\
\dot{Q}_\nu
&=&\frac{(kT)^4}{2\pi^2\hbar^3c^3}\int dx I_{\nu,0}\\
\frac{d{H}_\nu}{dt}&=&-\sigma_{\text{tr}} n_Ncx^2{H}_\nu
}
}

Again, comparing the anisotropic and isotropic results, we see that the anisotropic formula is the isotropic formula, augmented with terms of the form $H_\nu^2$ and $K_\nu^2$.  These are much smaller than the isotropic part. Therefore, isotropic simulations can capture the main properties very well. We henceforth focus on the isotropic case.

\subsection{Higher-Order Corrections}
\label{higher}
Now let's look back to the isotropic case.
There are three different kinds of residuals in that derivation. The $O(\frac{1}{\beta m})$ residuals come from the residuals in $a_2$, while the $O(\frac{x}{\beta m})$ and $O(\frac{x^2}{\beta m})$ residuals come from $a_3$ and $a_4$. If the neutrino energy is close to the nucleon temperature, i.e. $x=O(1)$, these residuals are the same order. However, $\frac{x^2}{\beta m}=\frac{\varepsilon^2}{mkT}$ can sometimes be larger than $0.5$, or even reach $1.0$, and these are not very small numbers. The Kompaneets equation itself that handles neutrino number redistribution is less affected by such residuals, because it automatically conserves the particle number $-$ these residuals will cancel out. However, in the formula for the energy deposition rate, such residuals will add up and the relative error of the total energy deposition rate can be greater than 20\% (see Table \ref{tab:iso_exp}). Therefore, we will want to include higher-order terms in $\varepsilon/mc^2$ when calculating the total energy deposition rate.

Using eqs. \ref{equ:ak2}, \ref{equ:ak3}, and \ref{equ:ak4}, we derive that
\wide{
\equ{
I_\nu
&=&\sigma_{\text{tr}} n_Nc\frac{kT}{mc^2}\left\{x^6g^{(1)}-\frac{kT}{ mc^2}\frac{8x(2-\delta_N)-(50-22\delta_N)}{3-\delta_N}x^6g^{(1)}\right\}\nonumber\\
& &+\sigma_{\text{tr}} n_Nc\frac{kT}{mc^2}\left\{\frac{kT}{mc^2}\frac{2-\delta_N}{3-\delta_N}\frac{d}{dx}[x^8\frac{dg^{(1)}}{dx}+x^8g^{(1)}]-\frac{kT}{mc^2}\frac{2(2-\delta_N)}{3-\delta_N}x^8f^3\frac{d}{dx}(\frac{g^{(1)}}{f^2})\right\}\, ,
\label{equ:kom_corr}
}
}
where $g^{(1)}=\frac{df}{dx}+f-f^2$. Since we are more interested in $\dot{Q}_\nu$, we can throw away all total derivative terms, since they will vanish after the integration. The term with $f^3\frac{d}{dx}(\frac{g^{(1)}}{f^2})$ is quadratic in $f$, and in most cases it is much smaller than the linear terms in $f$, so we can also drop it. Then, we find a simplified flux with a correction factor in it:
\equ{
\tilde{I}_\nu
&=&
\sigma_{\text{tr}} n_Nc\frac{kT}{mc^2}x^6(\frac{df}{dx}+f-f^2)(1-\lambda)\, ,
\label{equ:dE_corr}
}
where the correction factor $\lambda$ is
\equ{
\lambda
&=&\frac{kT}{ mc^2}\frac{8x(2-\delta_N)-(50-22\delta_N)}{3-\delta_N}
\label{equ:factor}
}
 
\begin{figure*}[t]
    \centering
    \includegraphics[width=0.48\textwidth]{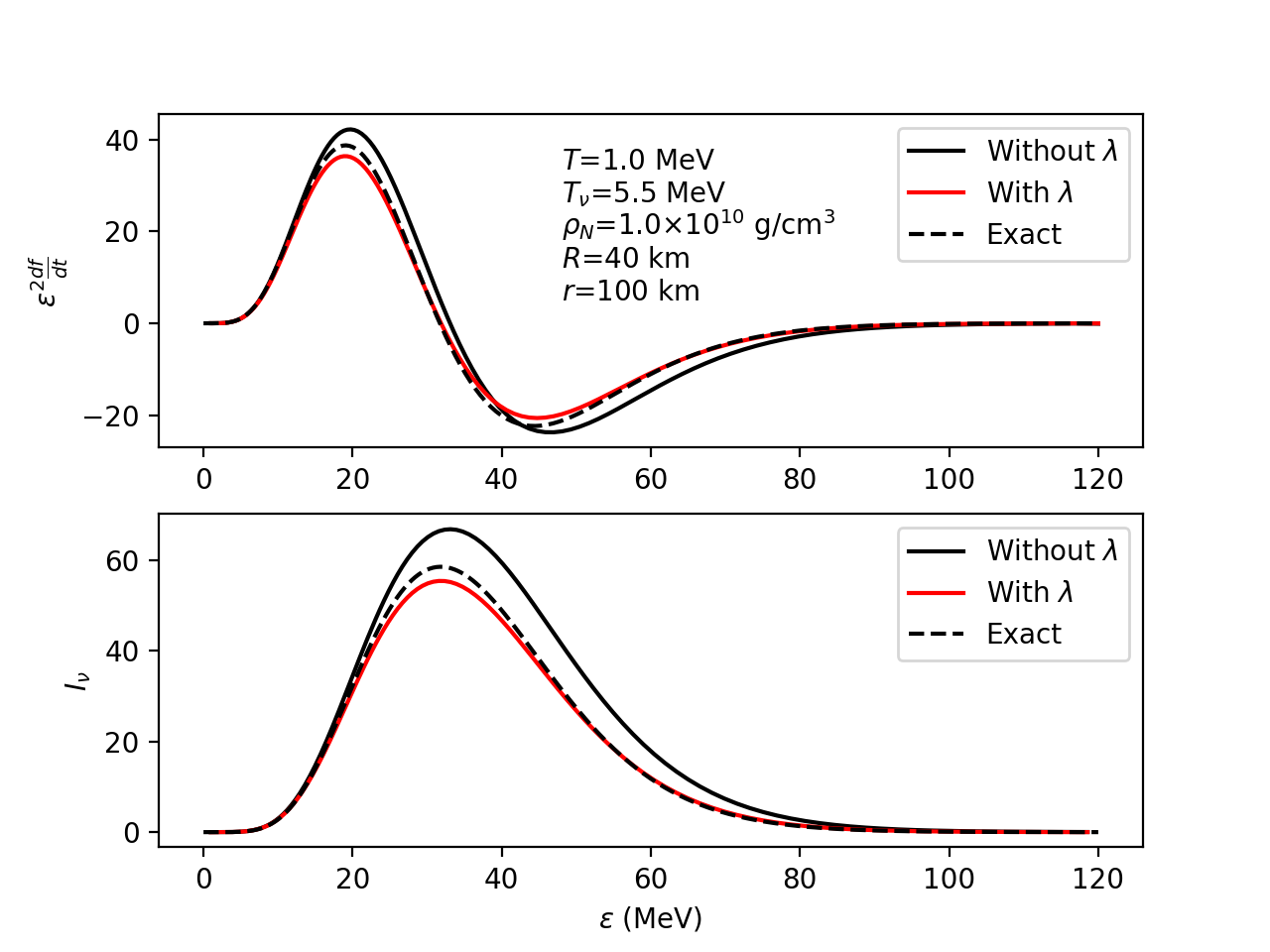}
    \includegraphics[width=0.48\textwidth]{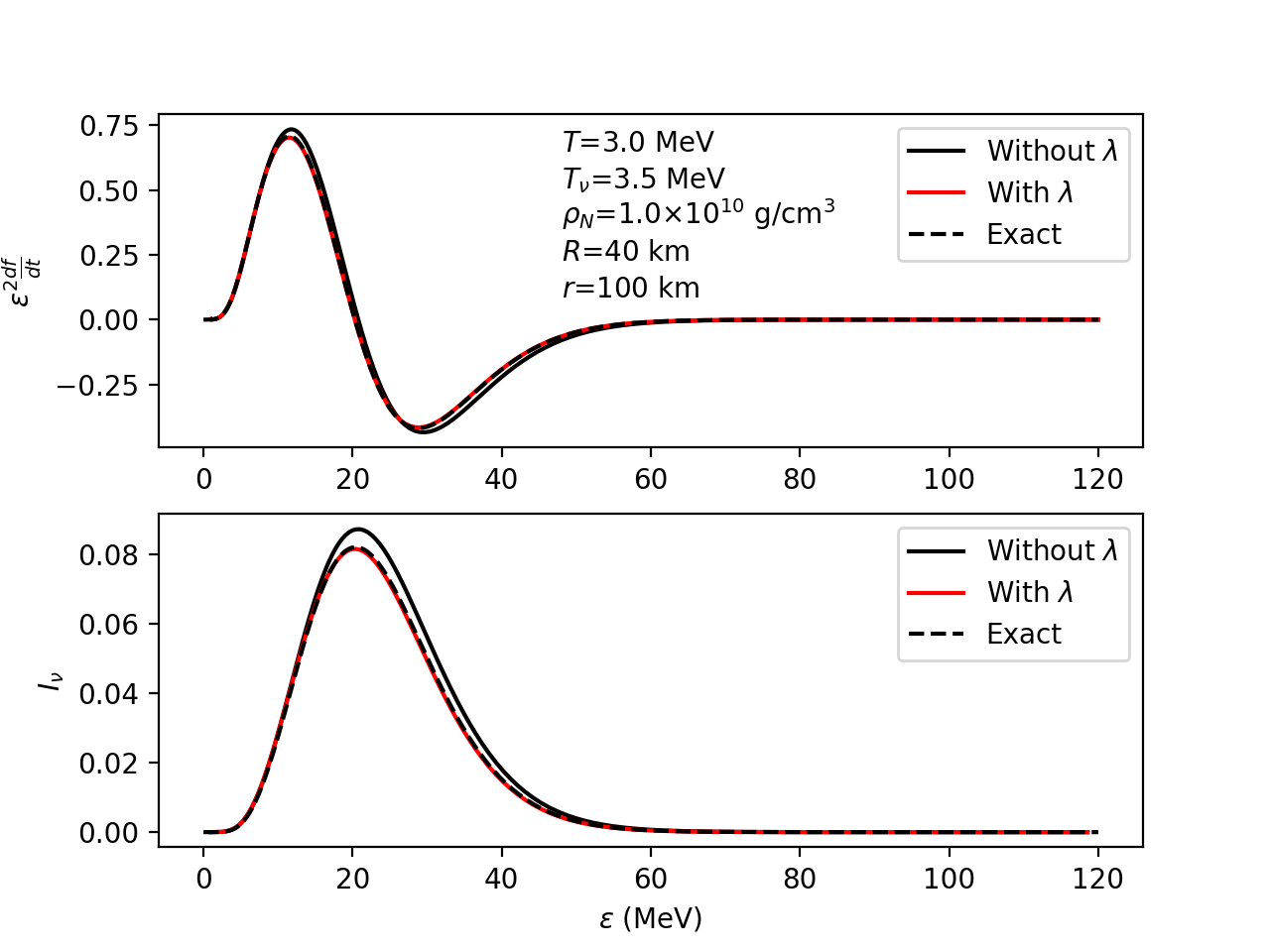}
    \caption{Comparison between particle number transfer and energy transfer calculated in different ways. Here, the nucleon temperature is $T=1$ MeV and neutrino temperature is $5.5$ MeV in the left two panels, while $T=3$ MeV and $T_\nu=3.5$ MeV in the right two panels. The top two panels show the particle number transfer calculated by the standard Kompaneets equation (black solid line), corrected Kompaneets equation (red solid line), and by doing the exact integration numerically (black dashed line). The bottom two panels show $I_\nu$ calculated with (red solid line) and without (black solid line) the correction factor. The exact energy deposition rate (black dashed line) is calculated by doing the integration numerically. The y-axes are arbitrarily scaled.}
    \label{fig:comp1}
\end{figure*}

$\tilde{I}_\nu$ is a good approximation for $I_\nu$. Each will give the same total energy deposition rate, and the errors in the spectrum evolution caused by the total derivative terms are quite small when the neutrino distribution has a Fermi-Dirac form, as shown in Figure \ref{fig:comp1}.

We do not need to derive expressions incorporating higher-order corrections for the anisotropic case.  $K_{\nu}$ at depth is zero, and stays smaller than $J_{\nu}$ in the important gain region. Concern $H_\nu$. Because we integrate out both $\theta$ and $\theta'$, the integrations involving $H_{\nu}$ and 
higher-order terms can only be non-zero if $\cos(\theta)$ and $\cos(\theta')$ are both of even order. But $H_\nu$ would pair with either one $\cos(\theta)$ or one $\cos(\theta')$ and an odd number of $H_\nu$s will introduce an odd number of $\cos(\theta)$s or $\cos(\theta')$s. Thus, all odd order $H_\nu$ terms vanish, and the lowest anisotropic contribution will be the quadratic term of $H_\nu$, which is negligible. As a result, we can safely ignore anisotropic terms, even when we introduce higher-order terms
in $\varepsilon/mc^2$.

In conclusion, the corrected Kompaneets equation and energy deposition rate for the isotropic (and, in practice, general) case are:
\equ{
\tilde{I}_\nu&=&\frac{kT}{mc^2}n_N\sigma_{\text{tr}}cx^6(\frac{df}{dx}+f-f^2)(1-\lambda)\\
\frac{df}{dt}
&=&\frac{1}{x^2}\frac{d\tilde{I}_\nu}{dx}\nonumber\\
&=&\frac{kT}{mc^2}n_N\sigma_{\text{tr}}c\frac{1}{x^2}\frac{d}{dx}\left[x^6(\frac{df}{dx}+f-f^2)(1-\lambda)\right]\\
\dot{q}_\nu
&=&kT\left(\tilde{I}_\nu-\frac{d}{dx}(x\tilde{I}_\nu)\right)\\
\dot{Q}_\nu
&=&\frac{(kT)^4}{2\pi^2\hbar^3c^3}\int dx\tilde{I}_\nu\, ,
}
where the correction factor is
\equ{
\lambda
&=&\frac{kT}{ mc^2}\frac{8x(2-\delta_N)-(50-22\delta_N)}{3-\delta_N}
}
These equations are valid for nucleon temperature $T\ll m$ and neutrino temperature $T_\nu\ll\sqrt{mkT}$. In anisotropic cases, if we ignore higher order correction of all quadratic terms in $f$, we can simply replace $\frac{df}{dt}$ and $\dot{q}_\nu$ in the above equations with $\frac{dJ_\nu}{dt}$ and $\dot{q}_{\nu,0}$ and add the anisotropic terms:
{
\equ{
\tilde{I}_{\nu,0}&=&\frac{kT}{mc^2}n_N\sigma_{\text{tr}}cx^6\bigg[(\frac{dJ_\nu}{dx}+J_\nu-J_\nu^2)(1-\lambda)\nonumber\\&&+\frac{3-3\delta_N}{3-\delta_N}H_\nu^2+\frac{2}{27}\frac{3\delta_N}{3-\delta_N}K_\nu^2\bigg]\\
\frac{dJ_\nu}{dt}
&=&\frac{1}{x^2}\frac{d\tilde{I}_{\nu,0}}{dx}\\
\dot{q}_{\nu,0}
&=&kT\left(\tilde{I}_{\nu,0}-\frac{d}{dx}(x\tilde{I}_{\nu,0})\right)\\
\dot{Q}_\nu
&=&\frac{(kT)^4}{2\pi^2\hbar^3c^3}\int dx\tilde{I}_{\nu,0}\, ,
}
}

As stated earlier, the total derivative form of the Kompaneets equation will automatically conserve the particle number and will have good numerical behavior. Including the higher-order energy correction term preserves this property. Since the correction can be describe by a factor which is independent of the distribution function, this scheme is easy to implement numerically.

\begin{figure*}[t]
    \centering
    \includegraphics[width=0.48\textwidth]{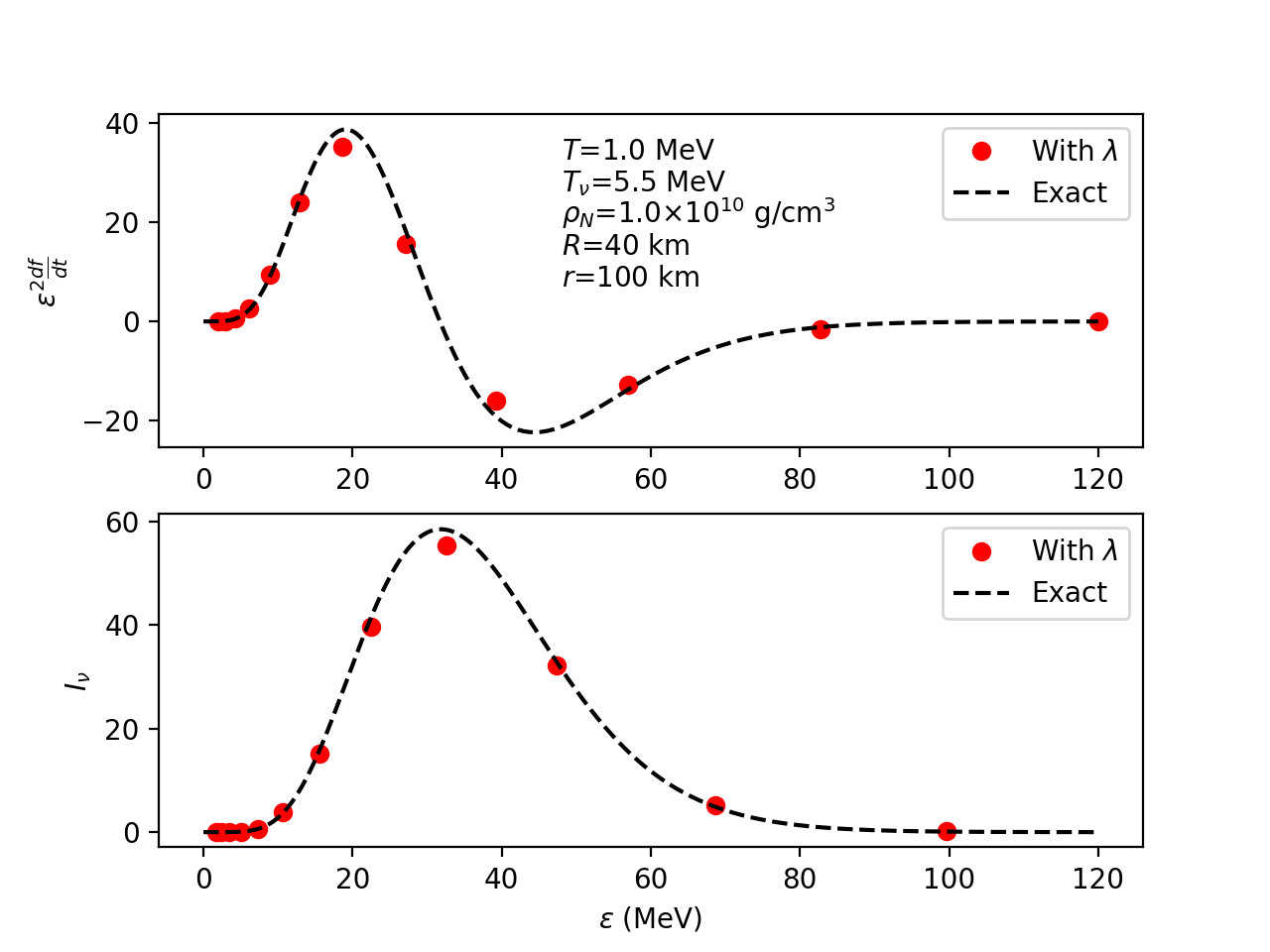}
    \includegraphics[width=0.48\textwidth]{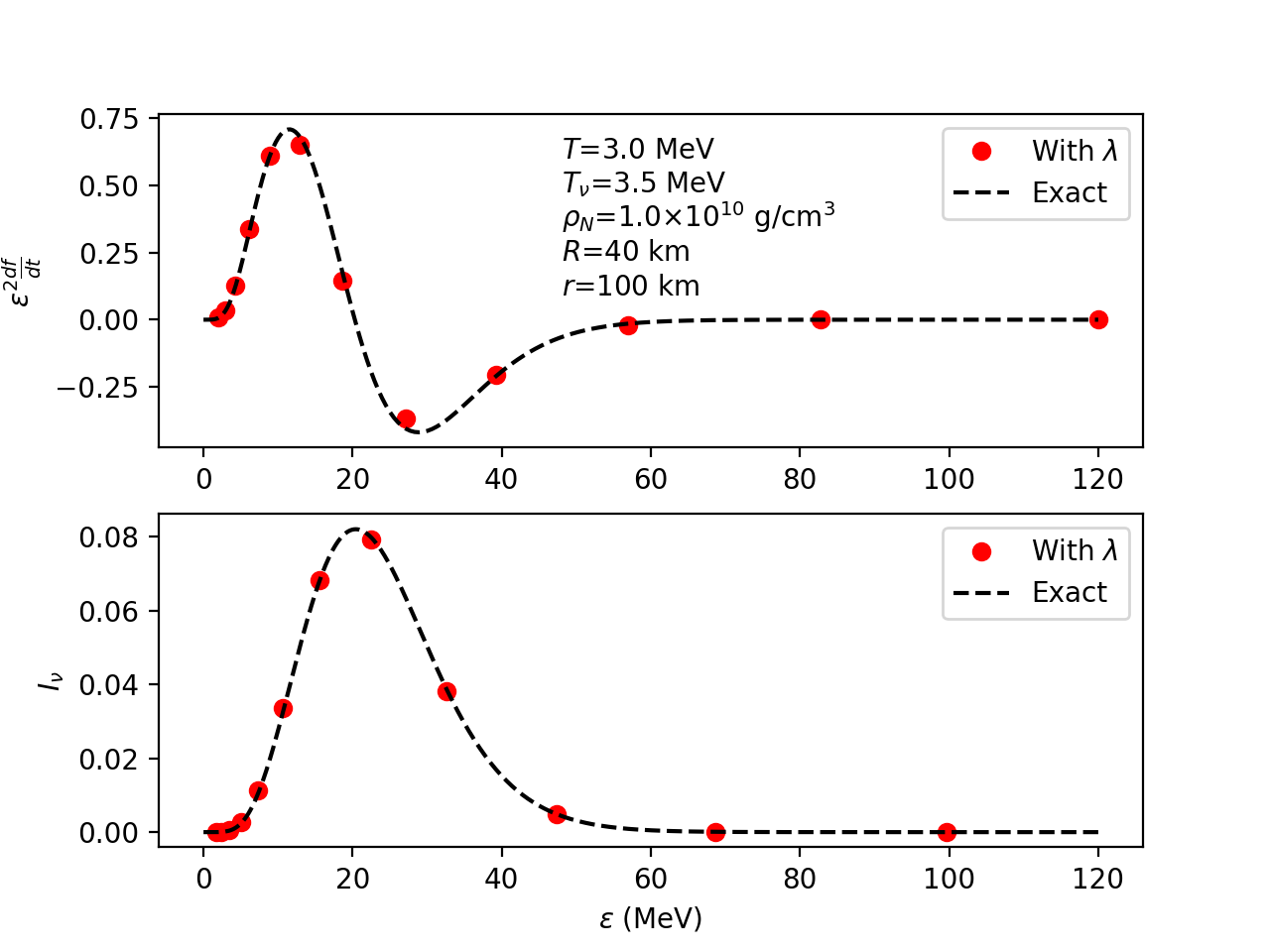}
   \caption{Comparison between particle number transfer and energy transfer calculated in different ways with only twelve energy bins. The nucleon temperature is $T=1$ MeV and neutrino temperature is $5.5$ MeV in left two panels, while $T=3$ MeV and $T_\nu=3.5$ MeV in the right two panels. The top two panels show the particle number transfer calculated by the corrected Kompaneets equation (red dots), and by doing the integration numerically (black dashed line). The bottom two panels depict $I_\nu$ calculated with the correction factor (red dots). The exact energy deposition rate (black dashed line) is calculated by doing the integration numerically. The y-axes are arbitrarily scaled. Red dots in the upper panels are located at the centers of each logarithmic energy bin, while in the lower panels they are located at the edges.}
    \label{fig:comp4}
\end{figure*}

\begin{table}[t]
    \caption{Relative errors of the total energy deposition rate under different parameter assumptions, i.e., $\delta=\frac{\dot{Q}_\nu-\dot{Q}_{\nu,e}}{\dot{Q}_{\nu,e}}$. $\dot{Q}_{\nu,e}$ is the exact energy deposition calculated by numerical integration. We set $R=40$ km and $r=100$ km here. For coarse bins, we use twelve logarithmic energy groups between 2 MeV and 120 MeV. Although the relative errors calculated by coarse binning are a bit smaller than the analytical errors, it just indicates that the numerical and analytical errors have different signs and that they partially cancel out. This behavior depends on the numerical algorithm, but one can see that numerical errors are small, even when there are only twelve energy bins. The unit of $T$ and $T_\nu$ is MeV.}
\begin{ruledtabular}
\begin{tabular}{ccccc}
$T$  & $T_\nu$  & Without $\lambda$ & With $\lambda$ & Coarse Bins \\
\colrule
1         & 3.5             & 13.1\%             & -1.0\%          & -0.7\%      \\
1         & 4.5             & 17.3\%             & -2.1\%          & -1.8\%      \\
1         & 5.5             & 21.6\%             & -3.4\%          & -3.3\%      \\
2         & 3.5             & 10.9\%             & -0.8\%          & -0.5\%      \\
2         & 4.5             & 15.2\%             & -1.8\%          & -1.5\%      \\
2         & 5.5             & 19.5\%             & -3.0\%          & -2.8\%      \\
3         & 3.5             & 8.8\%             & -0.6\%          & -0.2\%      \\
3         & 4.5             & 13.2\%             & -1.4\%          & -1.0\%      \\
3         & 5.5             & 17.5\%             & -2.4\%          & -2.3\%      \\
\end{tabular}
\end{ruledtabular}
    \label{tab:iso_exp}
\end{table}

\begin{table}[t]
    \caption{{Total energy deposition rate $\dot{Q}_\nu$ and the total energy deposition rate per unit mass $\dot{Q}_\nu/\rho_N$ under different parameter assumptions.} $\dot{Q}_{\nu,e}$ is the exact energy deposition calculated by numerical integration, while $\dot{Q}_{\nu,\lambda}$ is calculated using the formula with the correction factor. We set $R=40$ km and $r=100$ km here. Other parameters are the same as in Table \ref{tab:iso_exp}. Values in this table are calculated for only one neutrino specie, e.g. for $\mu$-neutrino. {$T$ and $T_\nu$ are in MeV, while the units for $\frac{\dot{Q}_{\nu}}{\rho_N}$ and $\dot{Q}_\nu$ are erg g$^{-1}$ s$^{-1}$ and erg cm$^3$s$^{-1}$, respectively.}}
\begin{ruledtabular}
\begin{tabular}{ccccc}
$T$  & $T_\nu$  & $\frac{\dot{Q}_{\nu,e}}{\rho_N}$ &$\frac{\dot{Q}_{\nu,\lambda}}{\rho_N}$ &$\dot{Q}_{\nu,\lambda}$ \\
\colrule
1         & 3.5             & $1.20\times10^{18}$           & $1.18\times10^{18}$       & $1.18\times10^{28}$\\
1         & 4.5             & $7.29\times10^{18}$           & $7.14\times10^{18}$       & $7.14\times10^{28}$\\
1         & 5.5             & $3.01\times10^{19}$           & $2.91\times10^{19}$        & $2.91\times10^{29}$     \\
2         & 3.5             & $7.36\times10^{17}$           & $7.30\times10^{17}$       & $7.30\times10^{27}$       \\
2         & 4.5             & $5.32\times10^{18}$           & $5.23\times10^{18}$       & $5.23\times10^{28}$       \\
2         & 5.5             & $2.39\times10^{19}$           & $2.32\times10^{19}$        & $2.32\times10^{29}$      \\
3         & 3.5             & $2.58\times10^{17}$           & $2.56\times10^{17}$       & $2.56\times10^{27}$        \\
3         & 4.5             & $3.27\times10^{18}$           & $3.23\times10^{18}$       & $3.23\times10^{28}$        \\
3         & 5.5             & $1.74\times10^{19}$           & $1.70\times10^{19}$       & $1.70\times10^{29}$        \\
\end{tabular}
\end{ruledtabular}
    \label{tab:values}
\end{table}

\begin{figure*}[t]
    \centering
    \includegraphics[width=0.48\textwidth]{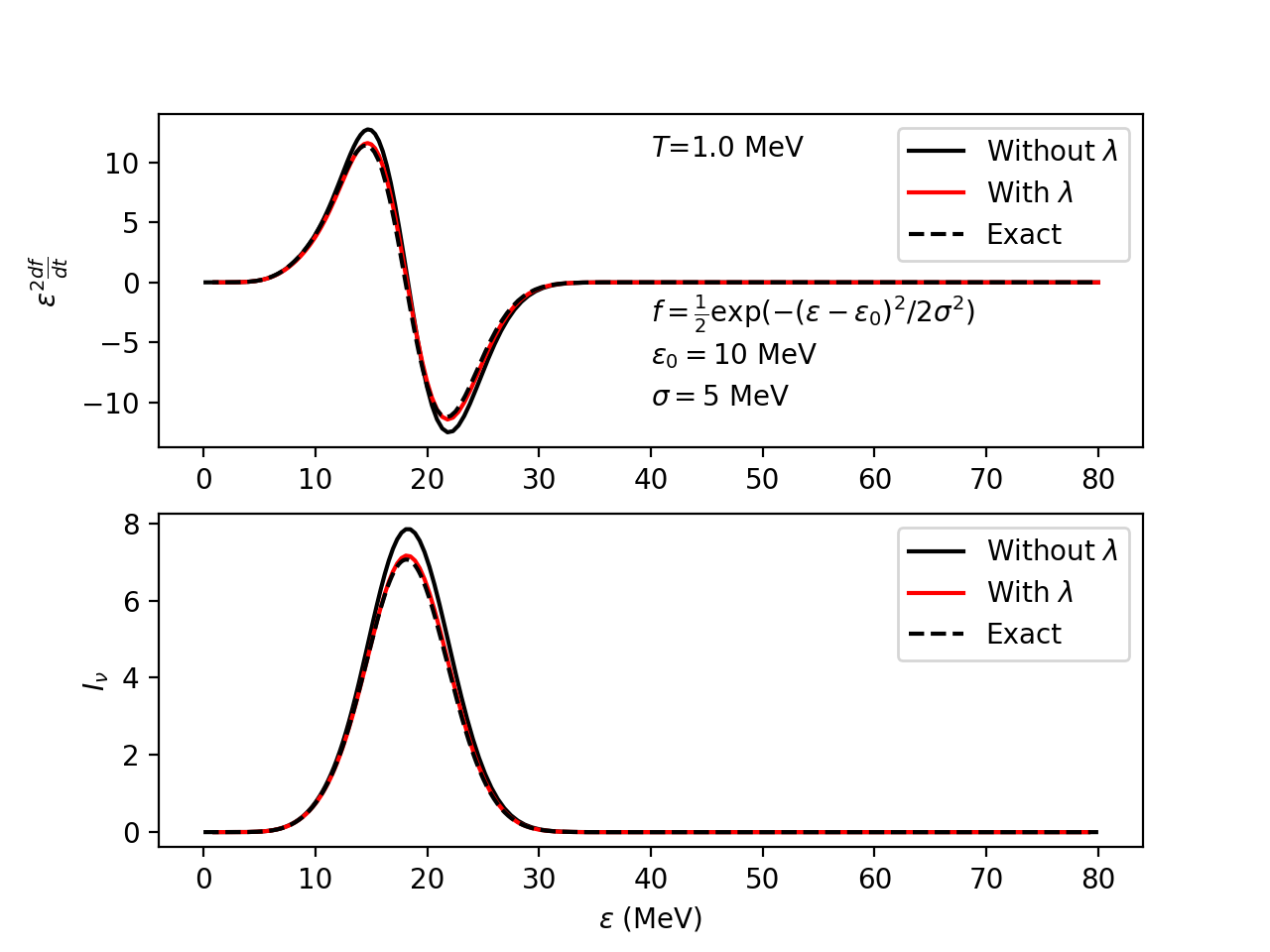}
    \includegraphics[width=0.48\textwidth]{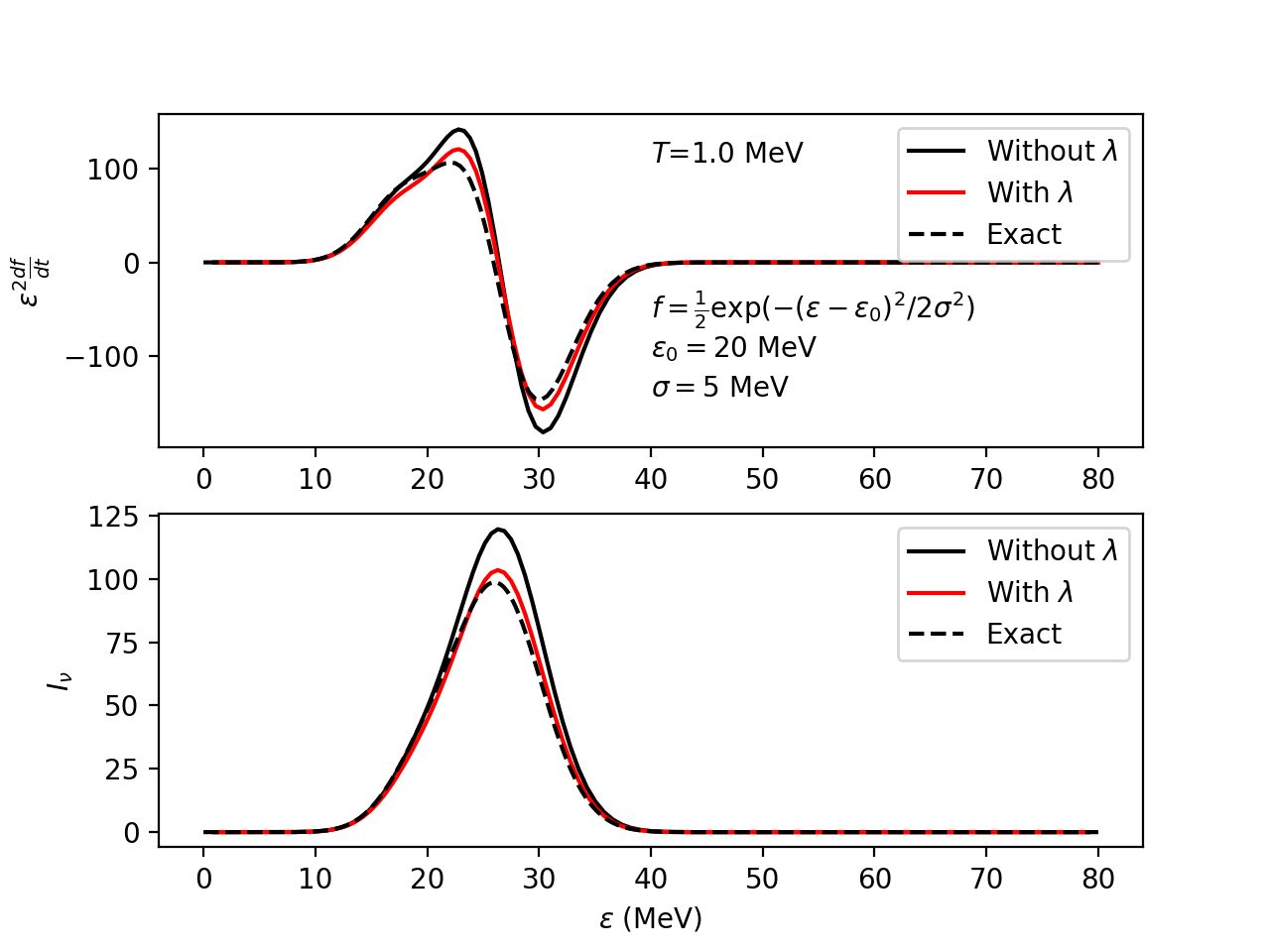}
    \caption{Comparison between particle number transfer and energy transfer calculated in different ways. The neutrino distribution is $f=\frac{1}{2}\exp(-\frac{(\varepsilon-\varepsilon_0)^2}{2\sigma^2})$. This distribution is very different from the real case, but the formulae work well. Left panels: $\varepsilon_0=10$ MeV and $\sigma=5$ MeV. Right panels: $\varepsilon_0=20$ MeV and $\sigma=5$ MeV. Top panels: Particle number transfer calculated by standard Kompaneets equation (black solid line), corrected Kompaneets equation (red solid line) and by doing the integration numerically (black dashed line). Bottom panels: $I_\nu$ calculated with (red solid line) and without (black solid line) the correction factor. The exact energy deposition rate (black dashed line) is calculated by doing the integration numerically. The y-axes are arbitrarily scaled. Note that red lines fluctuate around the exact values. This is because we throw away the total derivative terms in the correction. Such fluctuations have little effect on $\dot{Q}_\nu$, and the relative errors of $\dot{Q}_\nu$ in both cases are below 2\%. }
    \label{fig:gaussian}
\end{figure*}

\section{Numerical Tests}

In this section, we provide the results of some numerical tests. All numerical tests are done with isotropic neutrino distributions with a dilution factor: $f=\frac{1}{e^{\varepsilon/kT}+1}(1-\sqrt{1-\frac{R^2}{r^2}})$. Here, $R$ is the neutrinosphere radius and we set $R=40$ km, while $r$ is set to be 100 km. The nucleon density is set to be $10^{10}$ g/cm$^3$ so that the non-degenerate nucleon approximation (Eq. (\ref{equ:approx})) is valid. The electron fraction ($Y_e$) is set to  $0.2$. {The boundary conditions for the Kompaneets equation are set to be ``zero-flux'', i.e., no particles are scattered outside the energy region we consider.} 

Figure \ref{fig:comp1} shows the comparison between results with and without the correction factor $\lambda$. The neutrino distributions in these figures are Fermi-Dirac at temperature $T_\nu$. Table \ref{tab:iso_exp} shows the relative error of the total energy deposition rate $\dot{Q}_\nu$, with and without the correction factor. These are termed here ``analytical errors'', since they are caused by the residuals we ignore in our derivation. {Table \ref{tab:values} lists the values of $\dot{Q}$ and $\dot{Q}/\rho_N$. In our scheme, $\dot{Q}$ is proportional to $\rho_N$, so only the ratio is useful for testing numerical calculations when the Gaussian approximation (Eq. (\ref{equ:approx})) is employed.  However, a comparison of $\dot{Q}$ with and without this approximation can be used to check if Eq. (\ref{equ:approx}) works well. For the density range in the important gain region, it does.}

Our formulae work well even when the energy binning is coarse. Figure \ref{fig:comp4} uses the same models as Figures \ref{fig:comp1}, but the number of energy bins is taken to be only twelve. We use logarithmic energy binning between 2 MeV and 120 MeV, as stated in the figures. The final relative error are given in the last row of Table \ref{tab:iso_exp}. We see that the relative errors do not change much when we use coarse energy bins. This good numerical behavior partly comes from the property of the Kompaneets equation. We can calculate the particle number flux at the edge of each energy bin, so we can get the exact number of particles that enter and leave this energy bin. Thus, numerical accuracy weakly depends on the number of energy bins.

We've experimented with other neutrino distribution functions to test our formula. Figure \ref{fig:gaussian} is calculated with the neutrino distribution given by $f=\frac{1}{2}\exp(-\frac{(\varepsilon-\varepsilon_0)^2}{2\sigma^2})$, where $\sigma=5$ MeV and $\varepsilon_0$ are 10 and 20 MeV for the left and right panels. Such a distribution is very different from the realistic case, but even here the formulae work well. The relative errors in $\dot{Q}_\nu$ are below 2\%. From the slight deviations of the red lines from the exact values seen in the Figure \ref{fig:gaussian}, we see that the total derivative terms that we threw away have some effect, and this might influence the particle number redistribution in some bins. Therefore, if one wants to handle all possible functional forms, instead of focusing on Fermi-Dirac like functions, we suggest using the exact particle number flux (Eq. (\ref{equ:kom_corr})).

\section{Conclusions}
In the context of low-energy-transfer neutrino-nucleon inelastic scattering, 
we have derived a generalized Kompaneets equation to handle the associated neutrino number 
redistribution and the rate of neutrino-matter energy transfer. The latter is a key new result.  
Unlike many previous approaches, the solves are linear in number of energy groups and include the angular anisotropy of the neutrino field when updating the zeroth moment (spectral neutrino number density) of the radiation field. We have also derived the next higher-order correction in $\varepsilon/mc^2$ to the neutrino Kompaneets equation and matter heating rate.

A byproduct of this methodology is a straightforward formula for calculating the matter 
heating rate due to this process.  Neutrino-nucleon scattering is a subdominant 
heating process, but can have leverage when a CCSN model is marginally explodable 
in determining whether in fact a model explodes. In addition, when this process 
is included in sophisticated supernova codes the derived equation should facilitate the convergence
of the transport solutions for $\nu_{\mu}$, $\bar{\nu}_{\mu}$, $\nu_{\tau}$, and $\bar{\nu}_{\tau}$
neutrinos, for which the associated absorption processes that generally help stabilize
numerical solutions are weaker than for $\nu_e$ and $\bar{\nu}_e$ neutrinos.  

{To recap, our derivation of the Kompaneets formalism is based on the dynamic structure function for non-interactive nucleons, so the initial expression of the Boltzmann equation Eq. (\ref{equ:def}) automatically includes the recoil corrections calculated in \cite{horowitz2002}. However, with the Gaussian approximation Eq. (\ref{equ:approx}), we assume a low nucleon number density and throw away the phase space blocking term. In addition, our correction factor $\lambda$ includes thermal effects of the nucleons, not included in \cite{horowitz2002}. \citet{horowitz:17} considered in their static structure function the nucleon blocking term as well as the interactions between nucleons. However, since neither effect is important in the low number density regime we consider and which is relevant to the issue of the energy deposition rates in the gain region by inelastic neutrino-nucleon scattering, we do not include these effects here. In principle, our scheme can be generalized to include all these effects, but the lack of a satisfactory nucleon model (relevant only at high densities) remains an obstacle. The s-quark contribution considered in \cite{melson:15b} can be included in our result by simply changing the axial-vector coupling constant $g_A$. However, the magnitude of the strangeness correction these authors employed might be larger than experiments allow \cite{burrows:18,ahmed2012,green2017}.}

Progress in supernova theory has paralleled advances in particle physics, nuclear physics,
and the numerical arts.  Useful and robust algorithms have traditionally facilitated
this progress.  Our hope is that the equations and formalism we have derived here for neutrino-nucleon 
inelastic redistribution will be of use broadly in the disparate extant codes now addressing 
neutrino transfer and the mechanism and dynamics of core-collapse supernova explosions.

\section*{Acknowledgments}

The authors acknowledge discussions with Josh Dolence and David Vartanyan.
AB acknowledges support from the U.S. Department of Energy Office of Science and the Office
of Advanced Scientific Computing Research via the Scientific Discovery
through Advanced Computing (SciDAC4) program and Grant DE-SC0018297
(subaward 00009650) and support from the U.S. NSF under Grants AST-1714267 
and PHY-1804048 (the latter via the Max-Planck/Princeton Center (MPPC) for Plasma Physics). An award of computer time was provided by the INCITE program using resources of the Argonne Leadership Computing Facility, which is a DOE Office of Science User Facility supported under Contract DE-AC02-06CH11357.
Finally, the authors note that they employed computational resources provided by the TIGRESS high
performance computer center at Princeton University, which is jointly
supported by the Princeton Institute for Computational Science and
Engineering (PICSciE) and the Princeton University Office of Information
Technology.

\bibliography{sn}

\begin{thebibliography}{54}%
\makeatletter
\providecommand \@ifxundefined [1]{%
 \@ifx{#1\undefined}
}%
\providecommand \@ifnum [1]{%
 \ifnum #1\expandafter \@firstoftwo
 \else \expandafter \@secondoftwo
 \fi
}%
\providecommand \@ifx [1]{%
 \ifx #1\expandafter \@firstoftwo
 \else \expandafter \@secondoftwo
 \fi
}%
\providecommand \natexlab [1]{#1}%
\providecommand \enquote  [1]{``#1''}%
\providecommand \bibnamefont  [1]{#1}%
\providecommand \bibfnamefont [1]{#1}%
\providecommand \citenamefont [1]{#1}%
\providecommand \href@noop [0]{\@secondoftwo}%
\providecommand \href [0]{\begingroup \@sanitize@url \@href}%
\providecommand \@href[1]{\@@startlink{#1}\@@href}%
\providecommand \@@href[1]{\endgroup#1\@@endlink}%
\providecommand \@sanitize@url [0]{\catcode `\\12\catcode `\$12\catcode
  `\&12\catcode `\#12\catcode `\^12\catcode `\_12\catcode `\%12\relax}%
\providecommand \@@startlink[1]{}%
\providecommand \@@endlink[0]{}%
\providecommand \url  [0]{\begingroup\@sanitize@url \@url }%
\providecommand \@url [1]{\endgroup\@href {#1}{\urlprefix }}%
\providecommand \urlprefix  [0]{URL }%
\providecommand \Eprint [0]{\href }%
\providecommand \doibase [0]{http://dx.doi.org/}%
\providecommand \selectlanguage [0]{\@gobble}%
\providecommand \bibinfo  [0]{\@secondoftwo}%
\providecommand \bibfield  [0]{\@secondoftwo}%
\providecommand \translation [1]{[#1]}%
\providecommand \BibitemOpen [0]{}%
\providecommand \bibitemStop [0]{}%
\providecommand \bibitemNoStop [0]{.\EOS\space}%
\providecommand \EOS [0]{\spacefactor3000\relax}%
\providecommand \BibitemShut  [1]{\csname bibitem#1\endcsname}%
\let\auto@bib@innerbib\@empty
\bibitem [{\citenamefont {{Bethe}}\ and\ \citenamefont
  {{Wilson}}(1985)}]{1985ApJ...295...14B}%
  \BibitemOpen
  \bibfield  {author} {\bibinfo {author} {\bibfnamefont {H.~A.}\ \bibnamefont
  {{Bethe}}}\ and\ \bibinfo {author} {\bibfnamefont {J.~R.}\ \bibnamefont
  {{Wilson}}},\ }\href {\doibase 10.1086/163343} {\bibfield  {journal}
  {\bibinfo  {journal} {\apj}\ }\textbf {\bibinfo {volume} {295}},\ \bibinfo
  {pages} {14} (\bibinfo {year} {1985})}\BibitemShut {NoStop}%
\bibitem [{\citenamefont {Bethe}(1990)}]{Bethe:1990mw}%
  \BibitemOpen
  \bibfield  {author} {\bibinfo {author} {\bibfnamefont {H.~A.}\ \bibnamefont
  {Bethe}},\ }\href {\doibase 10.1103/RevModPhys.62.801} {\bibfield  {journal}
  {\bibinfo  {journal} {Rev. Mod. Phys.}\ }\textbf {\bibinfo {volume} {62}},\
  \bibinfo {pages} {801} (\bibinfo {year} {1990})}\BibitemShut {NoStop}%
\bibitem [{\citenamefont {{Melson}}\ \emph
  {et~al.}(2015{\natexlab{a}})\citenamefont {{Melson}}, \citenamefont
  {{Janka}},\ and\ \citenamefont {{Marek}}}]{melson:15a}%
  \BibitemOpen
  \bibfield  {author} {\bibinfo {author} {\bibfnamefont {T.}~\bibnamefont
  {{Melson}}}, \bibinfo {author} {\bibfnamefont {H.-T.}\ \bibnamefont
  {{Janka}}}, \ and\ \bibinfo {author} {\bibfnamefont {A.}~\bibnamefont
  {{Marek}}},\ }\href {\doibase 10.1088/2041-8205/801/2/L24} {\bibfield
  {journal} {\bibinfo  {journal} {\apjl}\ }\textbf {\bibinfo {volume} {801}},\
  \bibinfo {eid} {L24} (\bibinfo {year} {2015}{\natexlab{a}})},\ \Eprint
  {http://arxiv.org/abs/1501.01961} {arXiv:1501.01961 [astro-ph.SR]}
  \BibitemShut {NoStop}%
\bibitem [{\citenamefont {{Lentz}}\ \emph {et~al.}(2015)\citenamefont
  {{Lentz}}, \citenamefont {{Bruenn}}, \citenamefont {{Hix}}, \citenamefont
  {{Mezzacappa}}, \citenamefont {{Messer}}, \citenamefont {{Endeve}},
  \citenamefont {{Blondin}}, \citenamefont {{Harris}}, \citenamefont
  {{Marronetti}},\ and\ \citenamefont {{Yakunin}}}]{lentz:15}%
  \BibitemOpen
  \bibfield  {author} {\bibinfo {author} {\bibfnamefont {E.~J.}\ \bibnamefont
  {{Lentz}}}, \bibinfo {author} {\bibfnamefont {S.~W.}\ \bibnamefont
  {{Bruenn}}}, \bibinfo {author} {\bibfnamefont {W.~R.}\ \bibnamefont {{Hix}}},
  \bibinfo {author} {\bibfnamefont {A.}~\bibnamefont {{Mezzacappa}}}, \bibinfo
  {author} {\bibfnamefont {O.~E.~B.}\ \bibnamefont {{Messer}}}, \bibinfo
  {author} {\bibfnamefont {E.}~\bibnamefont {{Endeve}}}, \bibinfo {author}
  {\bibfnamefont {J.~M.}\ \bibnamefont {{Blondin}}}, \bibinfo {author}
  {\bibfnamefont {J.~A.}\ \bibnamefont {{Harris}}}, \bibinfo {author}
  {\bibfnamefont {P.}~\bibnamefont {{Marronetti}}}, \ and\ \bibinfo {author}
  {\bibfnamefont {K.~N.}\ \bibnamefont {{Yakunin}}},\ }\href {\doibase
  10.1088/2041-8205/807/2/L31} {\bibfield  {journal} {\bibinfo  {journal}
  {\apjl}\ }\textbf {\bibinfo {volume} {807}},\ \bibinfo {eid} {L31} (\bibinfo
  {year} {2015})},\ \Eprint {http://arxiv.org/abs/1505.05110} {arXiv:1505.05110
  [astro-ph.SR]} \BibitemShut {NoStop}%
\bibitem [{\citenamefont {{Takiwaki}}\ \emph {et~al.}(2016)\citenamefont
  {{Takiwaki}}, \citenamefont {{Kotake}},\ and\ \citenamefont
  {{Suwa}}}]{TaKoSu16}%
  \BibitemOpen
  \bibfield  {author} {\bibinfo {author} {\bibfnamefont {T.}~\bibnamefont
  {{Takiwaki}}}, \bibinfo {author} {\bibfnamefont {K.}~\bibnamefont
  {{Kotake}}}, \ and\ \bibinfo {author} {\bibfnamefont {Y.}~\bibnamefont
  {{Suwa}}},\ }\href {\doibase 10.1093/mnrasl/slw105} {\bibfield  {journal}
  {\bibinfo  {journal} {\mnras}\ }\textbf {\bibinfo {volume} {461}},\ \bibinfo
  {pages} {L112} (\bibinfo {year} {2016})},\ \Eprint
  {http://arxiv.org/abs/1602.06759} {arXiv:1602.06759 [astro-ph.HE]}
  \BibitemShut {NoStop}%
\bibitem [{\citenamefont {{Roberts}}\ \emph {et~al.}(2016)\citenamefont
  {{Roberts}}, \citenamefont {{Ott}}, \citenamefont {{Haas}}, \citenamefont
  {{O'Connor}}, \citenamefont {{Diener}},\ and\ \citenamefont
  {{Schnetter}}}]{roberts:16}%
  \BibitemOpen
  \bibfield  {author} {\bibinfo {author} {\bibfnamefont {L.~F.}\ \bibnamefont
  {{Roberts}}}, \bibinfo {author} {\bibfnamefont {C.~D.}\ \bibnamefont
  {{Ott}}}, \bibinfo {author} {\bibfnamefont {R.}~\bibnamefont {{Haas}}},
  \bibinfo {author} {\bibfnamefont {E.~P.}\ \bibnamefont {{O'Connor}}},
  \bibinfo {author} {\bibfnamefont {P.}~\bibnamefont {{Diener}}}, \ and\
  \bibinfo {author} {\bibfnamefont {E.}~\bibnamefont {{Schnetter}}},\ }\href
  {\doibase 10.3847/0004-637X/831/1/98} {\bibfield  {journal} {\bibinfo
  {journal} {\apj}\ }\textbf {\bibinfo {volume} {831}},\ \bibinfo {eid} {98}
  (\bibinfo {year} {2016})},\ \Eprint {http://arxiv.org/abs/1604.07848}
  {arXiv:1604.07848 [astro-ph.HE]} \BibitemShut {NoStop}%
\bibitem [{\citenamefont {{Bruenn}}\ \emph {et~al.}(2016)\citenamefont
  {{Bruenn}}, \citenamefont {{Lentz}}, \citenamefont {{Hix}}, \citenamefont
  {{Mezzacappa}}, \citenamefont {{Harris}}, \citenamefont {{Messer}},
  \citenamefont {{Endeve}}, \citenamefont {{Blondin}}, \citenamefont
  {{Chertkow}}, \citenamefont {{Lingerfelt}}, \citenamefont {{Marronetti}},\
  and\ \citenamefont {{Yakunin}}}]{2016ApJ...818..123B}%
  \BibitemOpen
  \bibfield  {author} {\bibinfo {author} {\bibfnamefont {S.~W.}\ \bibnamefont
  {{Bruenn}}}, \bibinfo {author} {\bibfnamefont {E.~J.}\ \bibnamefont
  {{Lentz}}}, \bibinfo {author} {\bibfnamefont {W.~R.}\ \bibnamefont {{Hix}}},
  \bibinfo {author} {\bibfnamefont {A.}~\bibnamefont {{Mezzacappa}}}, \bibinfo
  {author} {\bibfnamefont {J.~A.}\ \bibnamefont {{Harris}}}, \bibinfo {author}
  {\bibfnamefont {O.~E.~B.}\ \bibnamefont {{Messer}}}, \bibinfo {author}
  {\bibfnamefont {E.}~\bibnamefont {{Endeve}}}, \bibinfo {author}
  {\bibfnamefont {J.~M.}\ \bibnamefont {{Blondin}}}, \bibinfo {author}
  {\bibfnamefont {M.~A.}\ \bibnamefont {{Chertkow}}}, \bibinfo {author}
  {\bibfnamefont {E.~J.}\ \bibnamefont {{Lingerfelt}}}, \bibinfo {author}
  {\bibfnamefont {P.}~\bibnamefont {{Marronetti}}}, \ and\ \bibinfo {author}
  {\bibfnamefont {K.~N.}\ \bibnamefont {{Yakunin}}},\ }\href {\doibase
  10.3847/0004-637X/818/2/123} {\bibfield  {journal} {\bibinfo  {journal}
  {\apj}\ }\textbf {\bibinfo {volume} {818}},\ \bibinfo {eid} {123} (\bibinfo
  {year} {2016})},\ \Eprint {http://arxiv.org/abs/1409.5779} {arXiv:1409.5779
  [astro-ph.SR]} \BibitemShut {NoStop}%
\bibitem [{\citenamefont {{M{\"u}ller}}\ \emph
  {et~al.}(2017{\natexlab{a}})\citenamefont {{M{\"u}ller}}, \citenamefont
  {{Melson}}, \citenamefont {{Heger}},\ and\ \citenamefont
  {{Janka}}}]{MuMeHe17}%
  \BibitemOpen
  \bibfield  {author} {\bibinfo {author} {\bibfnamefont {B.}~\bibnamefont
  {{M{\"u}ller}}}, \bibinfo {author} {\bibfnamefont {T.}~\bibnamefont
  {{Melson}}}, \bibinfo {author} {\bibfnamefont {A.}~\bibnamefont {{Heger}}}, \
  and\ \bibinfo {author} {\bibfnamefont {H.-T.}\ \bibnamefont {{Janka}}},\
  }\href {\doibase 10.1093/mnras/stx1962} {\bibfield  {journal} {\bibinfo
  {journal} {\mnras}\ }\textbf {\bibinfo {volume} {472}},\ \bibinfo {pages}
  {491} (\bibinfo {year} {2017}{\natexlab{a}})},\ \Eprint
  {http://arxiv.org/abs/1705.00620} {arXiv:1705.00620 [astro-ph.SR]}
  \BibitemShut {NoStop}%
\bibitem [{\citenamefont {{O\'{}Connor}}\ and\ \citenamefont
  {{Couch}}(2018)}]{OcCo18}%
  \BibitemOpen
  \bibfield  {author} {\bibinfo {author} {\bibfnamefont {E.~P.}\ \bibnamefont
  {{O\'{}Connor}}}\ and\ \bibinfo {author} {\bibfnamefont {S.~M.}\ \bibnamefont
  {{Couch}}},\ }\href {\doibase 10.3847/1538-4357/aaa893} {\bibfield  {journal}
  {\bibinfo  {journal} {\apj}\ }\textbf {\bibinfo {volume} {854}},\ \bibinfo
  {eid} {63} (\bibinfo {year} {2018})}\BibitemShut {NoStop}%
\bibitem [{\citenamefont {{O'Connor}}\ and\ \citenamefont
  {{Couch}}(2018)}]{oconnor_couch2018}%
  \BibitemOpen
  \bibfield  {author} {\bibinfo {author} {\bibfnamefont {E.~P.}\ \bibnamefont
  {{O'Connor}}}\ and\ \bibinfo {author} {\bibfnamefont {S.~M.}\ \bibnamefont
  {{Couch}}},\ }\href {\doibase 10.3847/1538-4357/aadcf7} {\bibfield  {journal}
  {\bibinfo  {journal} {\apj}\ }\textbf {\bibinfo {volume} {865}},\ \bibinfo
  {eid} {81} (\bibinfo {year} {2018})},\ \Eprint
  {http://arxiv.org/abs/1807.07579} {arXiv:1807.07579 [astro-ph.HE]}
  \BibitemShut {NoStop}%
\bibitem [{\citenamefont {{Ott}}\ \emph {et~al.}(2018)\citenamefont {{Ott}},
  \citenamefont {{Roberts}}, \citenamefont {{da Silva Schneider}},
  \citenamefont {{Fedrow}}, \citenamefont {{Haas}},\ and\ \citenamefont
  {{Schnetter}}}]{ott2018_rel}%
  \BibitemOpen
  \bibfield  {author} {\bibinfo {author} {\bibfnamefont {C.~D.}\ \bibnamefont
  {{Ott}}}, \bibinfo {author} {\bibfnamefont {L.~F.}\ \bibnamefont
  {{Roberts}}}, \bibinfo {author} {\bibfnamefont {A.}~\bibnamefont {{da Silva
  Schneider}}}, \bibinfo {author} {\bibfnamefont {J.~M.}\ \bibnamefont
  {{Fedrow}}}, \bibinfo {author} {\bibfnamefont {R.}~\bibnamefont {{Haas}}}, \
  and\ \bibinfo {author} {\bibfnamefont {E.}~\bibnamefont {{Schnetter}}},\
  }\href {\doibase 10.3847/2041-8213/aaa967} {\bibfield  {journal} {\bibinfo
  {journal} {\apjl}\ }\textbf {\bibinfo {volume} {855}},\ \bibinfo {eid} {L3}
  (\bibinfo {year} {2018})}\BibitemShut {NoStop}%
\bibitem [{\citenamefont {{Burrows}}\ \emph {et~al.}(2019)\citenamefont
  {{Burrows}}, \citenamefont {{Radice}},\ and\ \citenamefont
  {{Vartanyan}}}]{burrows_low_2019}%
  \BibitemOpen
  \bibfield  {author} {\bibinfo {author} {\bibfnamefont {A.}~\bibnamefont
  {{Burrows}}}, \bibinfo {author} {\bibfnamefont {D.}~\bibnamefont {{Radice}}},
  \ and\ \bibinfo {author} {\bibfnamefont {D.}~\bibnamefont {{Vartanyan}}},\
  }\href {\doibase 10.1093/mnras/stz543} {\bibfield  {journal} {\bibinfo
  {journal} {\mnras}\ }\textbf {\bibinfo {volume} {485}},\ \bibinfo {pages}
  {3153} (\bibinfo {year} {2019})},\ \Eprint {http://arxiv.org/abs/1902.00547}
  {arXiv:1902.00547 [astro-ph.SR]} \BibitemShut {NoStop}%
\bibitem [{\citenamefont {{Vartanyan}}\ \emph {et~al.}(2019)\citenamefont
  {{Vartanyan}}, \citenamefont {{Burrows}}, \citenamefont {{Radice}},
  \citenamefont {{Skinner}},\ and\ \citenamefont {{Dolence}}}]{vartanyan2019}%
  \BibitemOpen
  \bibfield  {author} {\bibinfo {author} {\bibfnamefont {D.}~\bibnamefont
  {{Vartanyan}}}, \bibinfo {author} {\bibfnamefont {A.}~\bibnamefont
  {{Burrows}}}, \bibinfo {author} {\bibfnamefont {D.}~\bibnamefont {{Radice}}},
  \bibinfo {author} {\bibfnamefont {M.~A.}\ \bibnamefont {{Skinner}}}, \ and\
  \bibinfo {author} {\bibfnamefont {J.}~\bibnamefont {{Dolence}}},\ }\href@noop
  {} {\bibfield  {journal} {\bibinfo  {journal} {\mnras}\ }\textbf {\bibinfo
  {volume} {482}},\ \bibinfo {pages} {351} (\bibinfo {year}
  {2019})}\BibitemShut {NoStop}%
\bibitem [{\citenamefont {{Glas}}\ \emph {et~al.}(2019)\citenamefont {{Glas}},
  \citenamefont {{Just}}, \citenamefont {{Janka}},\ and\ \citenamefont
  {{Obergaulinger}}}]{2019ApJ...873...45G}%
  \BibitemOpen
  \bibfield  {author} {\bibinfo {author} {\bibfnamefont {R.}~\bibnamefont
  {{Glas}}}, \bibinfo {author} {\bibfnamefont {O.}~\bibnamefont {{Just}}},
  \bibinfo {author} {\bibfnamefont {H.~T.}\ \bibnamefont {{Janka}}}, \ and\
  \bibinfo {author} {\bibfnamefont {M.}~\bibnamefont {{Obergaulinger}}},\
  }\href {\doibase 10.3847/1538-4357/ab0423} {\bibfield  {journal} {\bibinfo
  {journal} {\apj}\ }\textbf {\bibinfo {volume} {873}},\ \bibinfo {eid} {45}
  (\bibinfo {year} {2019})},\ \Eprint {http://arxiv.org/abs/1809.10146}
  {arXiv:1809.10146 [astro-ph.HE]} \BibitemShut {NoStop}%
\bibitem [{\citenamefont {{Burrows}}\ \emph {et~al.}(2020)\citenamefont
  {{Burrows}}, \citenamefont {{Radice}}, \citenamefont {{Vartanyan}},
  \citenamefont {{Nagakura}}, \citenamefont {{Skinner}},\ and\ \citenamefont
  {{Dolence}}}]{2020MNRAS.491.2715B}%
  \BibitemOpen
  \bibfield  {author} {\bibinfo {author} {\bibfnamefont {A.}~\bibnamefont
  {{Burrows}}}, \bibinfo {author} {\bibfnamefont {D.}~\bibnamefont {{Radice}}},
  \bibinfo {author} {\bibfnamefont {D.}~\bibnamefont {{Vartanyan}}}, \bibinfo
  {author} {\bibfnamefont {H.}~\bibnamefont {{Nagakura}}}, \bibinfo {author}
  {\bibfnamefont {M.~A.}\ \bibnamefont {{Skinner}}}, \ and\ \bibinfo {author}
  {\bibfnamefont {J.~C.}\ \bibnamefont {{Dolence}}},\ }\href {\doibase
  10.1093/mnras/stz3223} {\bibfield  {journal} {\bibinfo  {journal} {\mnras}\
  }\textbf {\bibinfo {volume} {491}},\ \bibinfo {pages} {2715} (\bibinfo {year}
  {2020})},\ \Eprint {http://arxiv.org/abs/1909.04152} {arXiv:1909.04152
  [astro-ph.HE]} \BibitemShut {NoStop}%
\bibitem [{\citenamefont {{Burrows}}\ \emph {et~al.}(1995)\citenamefont
  {{Burrows}}, \citenamefont {{Hayes}},\ and\ \citenamefont
  {{Fryxell}}}]{bhf1995}%
  \BibitemOpen
  \bibfield  {author} {\bibinfo {author} {\bibfnamefont {A.}~\bibnamefont
  {{Burrows}}}, \bibinfo {author} {\bibfnamefont {J.}~\bibnamefont {{Hayes}}},
  \ and\ \bibinfo {author} {\bibfnamefont {B.~A.}\ \bibnamefont {{Fryxell}}},\
  }\href {\doibase 10.1086/176188} {\bibfield  {journal} {\bibinfo  {journal}
  {\apj}\ }\textbf {\bibinfo {volume} {450}},\ \bibinfo {pages} {830} (\bibinfo
  {year} {1995})},\ \Eprint {http://arxiv.org/abs/astro-ph/9506061}
  {astro-ph/9506061} \BibitemShut {NoStop}%
\bibitem [{\citenamefont {{Keil}}\ \emph {et~al.}(1995)\citenamefont {{Keil}},
  \citenamefont {{Janka}},\ and\ \citenamefont
  {{Raffelt}}}]{1995PhRvD..51.6635K}%
  \BibitemOpen
  \bibfield  {author} {\bibinfo {author} {\bibfnamefont {W.}~\bibnamefont
  {{Keil}}}, \bibinfo {author} {\bibfnamefont {H.~T.}\ \bibnamefont {{Janka}}},
  \ and\ \bibinfo {author} {\bibfnamefont {G.}~\bibnamefont {{Raffelt}}},\
  }\href {\doibase 10.1103/PhysRevD.51.6635} {\bibfield  {journal} {\bibinfo
  {journal} {\prd}\ }\textbf {\bibinfo {volume} {51}},\ \bibinfo {pages} {6635}
  (\bibinfo {year} {1995})},\ \Eprint {http://arxiv.org/abs/hep-ph/9410229}
  {arXiv:hep-ph/9410229 [hep-ph]} \BibitemShut {NoStop}%
\bibitem [{\citenamefont {{Burrows}}\ and\ \citenamefont
  {{Sawyer}}(1998)}]{1998PhRvC..58..554B}%
  \BibitemOpen
  \bibfield  {author} {\bibinfo {author} {\bibfnamefont {A.}~\bibnamefont
  {{Burrows}}}\ and\ \bibinfo {author} {\bibfnamefont {R.~F.}\ \bibnamefont
  {{Sawyer}}},\ }\href {\doibase 10.1103/PhysRevC.58.554} {\bibfield  {journal}
  {\bibinfo  {journal} {\prc}\ }\textbf {\bibinfo {volume} {58}},\ \bibinfo
  {pages} {554} (\bibinfo {year} {1998})},\ \Eprint
  {http://arxiv.org/abs/astro-ph/9801082} {astro-ph/9801082} \BibitemShut
  {NoStop}%
\bibitem [{\citenamefont {{Burrows}}\ and\ \citenamefont
  {{Sawyer}}(1999)}]{sawyer1999}%
  \BibitemOpen
  \bibfield  {author} {\bibinfo {author} {\bibfnamefont {A.}~\bibnamefont
  {{Burrows}}}\ and\ \bibinfo {author} {\bibfnamefont {R.~F.}\ \bibnamefont
  {{Sawyer}}},\ }\href {\doibase 10.1103/PhysRevC.59.510} {\bibfield  {journal}
  {\bibinfo  {journal} {\prc}\ }\textbf {\bibinfo {volume} {59}},\ \bibinfo
  {pages} {510} (\bibinfo {year} {1999})},\ \Eprint
  {http://arxiv.org/abs/astro-ph/9804264} {astro-ph/9804264} \BibitemShut
  {NoStop}%
\bibitem [{\citenamefont {{Roberts}}\ \emph {et~al.}(2012)\citenamefont
  {{Roberts}}, \citenamefont {{Reddy}},\ and\ \citenamefont
  {{Shen}}}]{roberts2012}%
  \BibitemOpen
  \bibfield  {author} {\bibinfo {author} {\bibfnamefont {L.~F.}\ \bibnamefont
  {{Roberts}}}, \bibinfo {author} {\bibfnamefont {S.}~\bibnamefont {{Reddy}}},
  \ and\ \bibinfo {author} {\bibfnamefont {G.}~\bibnamefont {{Shen}}},\ }\href
  {\doibase 10.1103/PhysRevC.86.065803} {\bibfield  {journal} {\bibinfo
  {journal} {\prc}\ }\textbf {\bibinfo {volume} {86}},\ \bibinfo {eid} {065803}
  (\bibinfo {year} {2012})},\ \Eprint {http://arxiv.org/abs/1205.4066}
  {arXiv:1205.4066 [astro-ph.HE]} \BibitemShut {NoStop}%
\bibitem [{\citenamefont {{Roberts}}\ and\ \citenamefont
  {{Reddy}}(2017)}]{roberts_reddy2017}%
  \BibitemOpen
  \bibfield  {author} {\bibinfo {author} {\bibfnamefont {L.~F.}\ \bibnamefont
  {{Roberts}}}\ and\ \bibinfo {author} {\bibfnamefont {S.}~\bibnamefont
  {{Reddy}}},\ }\href {\doibase 10.1103/PhysRevC.95.045807} {\bibfield
  {journal} {\bibinfo  {journal} {\prc}\ }\textbf {\bibinfo {volume} {95}},\
  \bibinfo {eid} {045807} (\bibinfo {year} {2017})},\ \Eprint
  {http://arxiv.org/abs/1612.02764} {arXiv:1612.02764 [astro-ph.HE]}
  \BibitemShut {NoStop}%
\bibitem [{\citenamefont {{Horowitz}}\ \emph {et~al.}(2017)\citenamefont
  {{Horowitz}}, \citenamefont {{Caballero}}, \citenamefont {{Lin}},
  \citenamefont {{O'Connor}},\ and\ \citenamefont {{Schwenk}}}]{horowitz:17}%
  \BibitemOpen
  \bibfield  {author} {\bibinfo {author} {\bibfnamefont {C.~J.}\ \bibnamefont
  {{Horowitz}}}, \bibinfo {author} {\bibfnamefont {O.~L.}\ \bibnamefont
  {{Caballero}}}, \bibinfo {author} {\bibfnamefont {Z.}~\bibnamefont {{Lin}}},
  \bibinfo {author} {\bibfnamefont {E.}~\bibnamefont {{O'Connor}}}, \ and\
  \bibinfo {author} {\bibfnamefont {A.}~\bibnamefont {{Schwenk}}},\ }\href
  {\doibase 10.1103/PhysRevC.95.025801} {\bibfield  {journal} {\bibinfo
  {journal} {\prc}\ }\textbf {\bibinfo {volume} {95}},\ \bibinfo {eid} {025801}
  (\bibinfo {year} {2017})},\ \Eprint {http://arxiv.org/abs/1611.05140}
  {arXiv:1611.05140 [nucl-th]} \BibitemShut {NoStop}%
\bibitem [{\citenamefont {{Burrows}}\ \emph {et~al.}(2018)\citenamefont
  {{Burrows}}, \citenamefont {{Vartanyan}}, \citenamefont {{Dolence}},
  \citenamefont {{Skinner}},\ and\ \citenamefont {{Radice}}}]{burrows:18}%
  \BibitemOpen
  \bibfield  {author} {\bibinfo {author} {\bibfnamefont {A.}~\bibnamefont
  {{Burrows}}}, \bibinfo {author} {\bibfnamefont {D.}~\bibnamefont
  {{Vartanyan}}}, \bibinfo {author} {\bibfnamefont {J.~C.}\ \bibnamefont
  {{Dolence}}}, \bibinfo {author} {\bibfnamefont {M.~A.}\ \bibnamefont
  {{Skinner}}}, \ and\ \bibinfo {author} {\bibfnamefont {D.}~\bibnamefont
  {{Radice}}},\ }\href {\doibase 10.1007/s11214-017-0450-9} {\bibfield
  {journal} {\bibinfo  {journal} {\ssr}\ }\textbf {\bibinfo {volume} {214}},\
  \bibinfo {eid} {33} (\bibinfo {year} {2018})}\BibitemShut {NoStop}%
\bibitem [{\citenamefont {{M{\"u}ller}}\ and\ \citenamefont
  {{Janka}}(2015)}]{muller_janka_pert}%
  \BibitemOpen
  \bibfield  {author} {\bibinfo {author} {\bibfnamefont {B.}~\bibnamefont
  {{M{\"u}ller}}}\ and\ \bibinfo {author} {\bibfnamefont {H.-T.}\ \bibnamefont
  {{Janka}}},\ }\href {\doibase 10.1093/mnras/stv101} {\bibfield  {journal}
  {\bibinfo  {journal} {\mnras}\ }\textbf {\bibinfo {volume} {448}},\ \bibinfo
  {pages} {2141} (\bibinfo {year} {2015})},\ \Eprint
  {http://arxiv.org/abs/1409.4783} {arXiv:1409.4783 [astro-ph.SR]} \BibitemShut
  {NoStop}%
\bibitem [{\citenamefont {{Couch}}\ \emph {et~al.}(2015)\citenamefont
  {{Couch}}, \citenamefont {{Chatzopoulos}}, \citenamefont {{Arnett}},\ and\
  \citenamefont {{Timmes}}}]{CoChAr15}%
  \BibitemOpen
  \bibfield  {author} {\bibinfo {author} {\bibfnamefont {S.~M.}\ \bibnamefont
  {{Couch}}}, \bibinfo {author} {\bibfnamefont {E.}~\bibnamefont
  {{Chatzopoulos}}}, \bibinfo {author} {\bibfnamefont {W.~D.}\ \bibnamefont
  {{Arnett}}}, \ and\ \bibinfo {author} {\bibfnamefont {F.~X.}\ \bibnamefont
  {{Timmes}}},\ }\href {\doibase 10.1088/2041-8205/808/1/L21} {\bibfield
  {journal} {\bibinfo  {journal} {\apjl}\ }\textbf {\bibinfo {volume} {808}},\
  \bibinfo {eid} {L21} (\bibinfo {year} {2015})},\ \Eprint
  {http://arxiv.org/abs/1503.02199} {arXiv:1503.02199 [astro-ph.HE]}
  \BibitemShut {NoStop}%
\bibitem [{\citenamefont {{Jones}}\ \emph {et~al.}(2016)\citenamefont
  {{Jones}}, \citenamefont {{R{\"o}pke}}, \citenamefont {{Pakmor}},
  \citenamefont {{Seitenzahl}}, \citenamefont {{Ohlmann}},\ and\ \citenamefont
  {{Edelmann}}}]{jones_2016}%
  \BibitemOpen
  \bibfield  {author} {\bibinfo {author} {\bibfnamefont {S.}~\bibnamefont
  {{Jones}}}, \bibinfo {author} {\bibfnamefont {F.~K.}\ \bibnamefont
  {{R{\"o}pke}}}, \bibinfo {author} {\bibfnamefont {R.}~\bibnamefont
  {{Pakmor}}}, \bibinfo {author} {\bibfnamefont {I.~R.}\ \bibnamefont
  {{Seitenzahl}}}, \bibinfo {author} {\bibfnamefont {S.~T.}\ \bibnamefont
  {{Ohlmann}}}, \ and\ \bibinfo {author} {\bibfnamefont {P.~V.~F.}\
  \bibnamefont {{Edelmann}}},\ }\href {\doibase 10.1051/0004-6361/201628321}
  {\bibfield  {journal} {\bibinfo  {journal} {\aap}\ }\textbf {\bibinfo
  {volume} {593}},\ \bibinfo {eid} {A72} (\bibinfo {year} {2016})},\ \Eprint
  {http://arxiv.org/abs/1602.05771} {arXiv:1602.05771 [astro-ph.SR]}
  \BibitemShut {NoStop}%
\bibitem [{\citenamefont {{Chatzopoulos}}\ \emph {et~al.}(2016)\citenamefont
  {{Chatzopoulos}}, \citenamefont {{Couch}}, \citenamefont {{Arnett}},\ and\
  \citenamefont {{Timmes}}}]{2016ApJ...822...61C}%
  \BibitemOpen
  \bibfield  {author} {\bibinfo {author} {\bibfnamefont {E.}~\bibnamefont
  {{Chatzopoulos}}}, \bibinfo {author} {\bibfnamefont {S.~M.}\ \bibnamefont
  {{Couch}}}, \bibinfo {author} {\bibfnamefont {W.~D.}\ \bibnamefont
  {{Arnett}}}, \ and\ \bibinfo {author} {\bibfnamefont {F.~X.}\ \bibnamefont
  {{Timmes}}},\ }\href {\doibase 10.3847/0004-637X/822/2/61} {\bibfield
  {journal} {\bibinfo  {journal} {\apj}\ }\textbf {\bibinfo {volume} {822}},\
  \bibinfo {eid} {61} (\bibinfo {year} {2016})},\ \Eprint
  {http://arxiv.org/abs/1601.05816} {arXiv:1601.05816 [astro-ph.SR]}
  \BibitemShut {NoStop}%
\bibitem [{\citenamefont {{M{\"u}ller}}\ \emph
  {et~al.}(2017{\natexlab{b}})\citenamefont {{M{\"u}ller}}, \citenamefont
  {{Melson}}, \citenamefont {{Heger}},\ and\ \citenamefont
  {{Janka}}}]{muller2017}%
  \BibitemOpen
  \bibfield  {author} {\bibinfo {author} {\bibfnamefont {B.}~\bibnamefont
  {{M{\"u}ller}}}, \bibinfo {author} {\bibfnamefont {T.}~\bibnamefont
  {{Melson}}}, \bibinfo {author} {\bibfnamefont {A.}~\bibnamefont {{Heger}}}, \
  and\ \bibinfo {author} {\bibfnamefont {H.-T.}\ \bibnamefont {{Janka}}},\
  }\href {\doibase 10.1093/mnras/stx1962} {\bibfield  {journal} {\bibinfo
  {journal} {\mnras}\ }\textbf {\bibinfo {volume} {472}},\ \bibinfo {pages}
  {491} (\bibinfo {year} {2017}{\natexlab{b}})},\ \Eprint
  {http://arxiv.org/abs/1705.00620} {arXiv:1705.00620 [astro-ph.SR]}
  \BibitemShut {NoStop}%
\bibitem [{\citenamefont {{Steiner}}\ \emph {et~al.}(2013)\citenamefont
  {{Steiner}}, \citenamefont {{Hempel}},\ and\ \citenamefont
  {{Fischer}}}]{steiner:13}%
  \BibitemOpen
  \bibfield  {author} {\bibinfo {author} {\bibfnamefont {A.~W.}\ \bibnamefont
  {{Steiner}}}, \bibinfo {author} {\bibfnamefont {M.}~\bibnamefont {{Hempel}}},
  \ and\ \bibinfo {author} {\bibfnamefont {T.}~\bibnamefont {{Fischer}}},\
  }\href {\doibase 10.1088/0004-637X/774/1/17} {\bibfield  {journal} {\bibinfo
  {journal} {\apj}\ }\textbf {\bibinfo {volume} {774}},\ \bibinfo {eid} {17}
  (\bibinfo {year} {2013})},\ \Eprint {http://arxiv.org/abs/1207.2184}
  {arXiv:1207.2184 [astro-ph.SR]} \BibitemShut {NoStop}%
\bibitem [{\citenamefont {{Schneider}}\ \emph {et~al.}(2019)\citenamefont
  {{Schneider}}, \citenamefont {{Roberts}}, \citenamefont {{Ott}},\ and\
  \citenamefont {{O'connor}}}]{schneider_eos_2019}%
  \BibitemOpen
  \bibfield  {author} {\bibinfo {author} {\bibfnamefont {A.~S.}\ \bibnamefont
  {{Schneider}}}, \bibinfo {author} {\bibfnamefont {L.~F.}\ \bibnamefont
  {{Roberts}}}, \bibinfo {author} {\bibfnamefont {C.~D.}\ \bibnamefont
  {{Ott}}}, \ and\ \bibinfo {author} {\bibfnamefont {E.}~\bibnamefont
  {{O'connor}}},\ }\href@noop {} {\bibfield  {journal} {\bibinfo  {journal}
  {arXiv e-prints}\ } (\bibinfo {year} {2019})},\ \Eprint
  {http://arxiv.org/abs/1906.02009} {arXiv:1906.02009 [astro-ph.HE]}
  \BibitemShut {NoStop}%
\bibitem [{\citenamefont {{Summa}}\ \emph {et~al.}(2018)\citenamefont
  {{Summa}}, \citenamefont {{Janka}}, \citenamefont {{Melson}},\ and\
  \citenamefont {{Marek}}}]{summa2018}%
  \BibitemOpen
  \bibfield  {author} {\bibinfo {author} {\bibfnamefont {A.}~\bibnamefont
  {{Summa}}}, \bibinfo {author} {\bibfnamefont {H.-T.}\ \bibnamefont
  {{Janka}}}, \bibinfo {author} {\bibfnamefont {T.}~\bibnamefont {{Melson}}}, \
  and\ \bibinfo {author} {\bibfnamefont {A.}~\bibnamefont {{Marek}}},\ }\href
  {\doibase 10.3847/1538-4357/aa9ce8} {\bibfield  {journal} {\bibinfo
  {journal} {\apj}\ }\textbf {\bibinfo {volume} {852}},\ \bibinfo {eid} {28}
  (\bibinfo {year} {2018})},\ \Eprint {http://arxiv.org/abs/1708.04154}
  {arXiv:1708.04154 [astro-ph.HE]} \BibitemShut {NoStop}%
\bibitem [{\citenamefont {{Burrows}}\ \emph {et~al.}(2007)\citenamefont
  {{Burrows}}, \citenamefont {{Dessart}}, \citenamefont {{Livne}},
  \citenamefont {{Ott}},\ and\ \citenamefont {{Murphy}}}]{burrows2007_mag}%
  \BibitemOpen
  \bibfield  {author} {\bibinfo {author} {\bibfnamefont {A.}~\bibnamefont
  {{Burrows}}}, \bibinfo {author} {\bibfnamefont {L.}~\bibnamefont
  {{Dessart}}}, \bibinfo {author} {\bibfnamefont {E.}~\bibnamefont {{Livne}}},
  \bibinfo {author} {\bibfnamefont {C.~D.}\ \bibnamefont {{Ott}}}, \ and\
  \bibinfo {author} {\bibfnamefont {J.}~\bibnamefont {{Murphy}}},\ }\href
  {\doibase 10.1086/519161} {\bibfield  {journal} {\bibinfo  {journal} {\apj}\
  }\textbf {\bibinfo {volume} {664}},\ \bibinfo {pages} {416} (\bibinfo {year}
  {2007})},\ \Eprint {http://arxiv.org/abs/astro-ph/0702539} {astro-ph/0702539}
  \BibitemShut {NoStop}%
\bibitem [{\citenamefont {{M{\"o}sta}}\ \emph {et~al.}(2014)\citenamefont
  {{M{\"o}sta}}, \citenamefont {{Richers}}, \citenamefont {{Ott}},
  \citenamefont {{Haas}}, \citenamefont {{Piro}}, \citenamefont {{Boydstun}},
  \citenamefont {{Abdikamalov}}, \citenamefont {{Reisswig}},\ and\
  \citenamefont {{Schnetter}}}]{MoRiOt14}%
  \BibitemOpen
  \bibfield  {author} {\bibinfo {author} {\bibfnamefont {P.}~\bibnamefont
  {{M{\"o}sta}}}, \bibinfo {author} {\bibfnamefont {S.}~\bibnamefont
  {{Richers}}}, \bibinfo {author} {\bibfnamefont {C.~D.}\ \bibnamefont
  {{Ott}}}, \bibinfo {author} {\bibfnamefont {R.}~\bibnamefont {{Haas}}},
  \bibinfo {author} {\bibfnamefont {A.~L.}\ \bibnamefont {{Piro}}}, \bibinfo
  {author} {\bibfnamefont {K.}~\bibnamefont {{Boydstun}}}, \bibinfo {author}
  {\bibfnamefont {E.}~\bibnamefont {{Abdikamalov}}}, \bibinfo {author}
  {\bibfnamefont {C.}~\bibnamefont {{Reisswig}}}, \ and\ \bibinfo {author}
  {\bibfnamefont {E.}~\bibnamefont {{Schnetter}}},\ }\href {\doibase
  10.1088/2041-8205/785/2/L29} {\bibfield  {journal} {\bibinfo  {journal}
  {\apjl}\ }\textbf {\bibinfo {volume} {785}},\ \bibinfo {eid} {L29} (\bibinfo
  {year} {2014})},\ \Eprint {http://arxiv.org/abs/1403.1230} {arXiv:1403.1230
  [astro-ph.HE]} \BibitemShut {NoStop}%
\bibitem [{\citenamefont {{Kuroda}}\ \emph {et~al.}(2020)\citenamefont
  {{Kuroda}}, \citenamefont {{Arcones}}, \citenamefont {{Takiwaki}},\ and\
  \citenamefont {{Kotake}}}]{2020arXiv200302004K}%
  \BibitemOpen
  \bibfield  {author} {\bibinfo {author} {\bibfnamefont {T.}~\bibnamefont
  {{Kuroda}}}, \bibinfo {author} {\bibfnamefont {A.}~\bibnamefont {{Arcones}}},
  \bibinfo {author} {\bibfnamefont {T.}~\bibnamefont {{Takiwaki}}}, \ and\
  \bibinfo {author} {\bibfnamefont {K.}~\bibnamefont {{Kotake}}},\ }\href@noop
  {} {\bibfield  {journal} {\bibinfo  {journal} {arXiv e-prints}\ ,\ \bibinfo
  {eid} {arXiv:2003.02004}} (\bibinfo {year} {2020})},\ \Eprint
  {http://arxiv.org/abs/2003.02004} {arXiv:2003.02004 [astro-ph.HE]}
  \BibitemShut {NoStop}%
\bibitem [{\citenamefont {{Burrows}}\ \emph {et~al.}(2012)\citenamefont
  {{Burrows}}, \citenamefont {{Dolence}},\ and\ \citenamefont
  {{Murphy}}}]{burrows2012}%
  \BibitemOpen
  \bibfield  {author} {\bibinfo {author} {\bibfnamefont {A.}~\bibnamefont
  {{Burrows}}}, \bibinfo {author} {\bibfnamefont {J.~C.}\ \bibnamefont
  {{Dolence}}}, \ and\ \bibinfo {author} {\bibfnamefont {J.~W.}\ \bibnamefont
  {{Murphy}}},\ }\href {\doibase 10.1088/0004-637X/759/1/5} {\bibfield
  {journal} {\bibinfo  {journal} {\apj}\ }\textbf {\bibinfo {volume} {759}},\
  \bibinfo {eid} {5} (\bibinfo {year} {2012})},\ \Eprint
  {http://arxiv.org/abs/1204.3088} {arXiv:1204.3088 [astro-ph.SR]} \BibitemShut
  {NoStop}%
\bibitem [{\citenamefont {{Blondin}}\ \emph {et~al.}(2003)\citenamefont
  {{Blondin}}, \citenamefont {{Mezzacappa}},\ and\ \citenamefont
  {{DeMarino}}}]{blondin2003}%
  \BibitemOpen
  \bibfield  {author} {\bibinfo {author} {\bibfnamefont {J.~M.}\ \bibnamefont
  {{Blondin}}}, \bibinfo {author} {\bibfnamefont {A.}~\bibnamefont
  {{Mezzacappa}}}, \ and\ \bibinfo {author} {\bibfnamefont {C.}~\bibnamefont
  {{DeMarino}}},\ }\href {\doibase 10.1086/345812} {\bibfield  {journal}
  {\bibinfo  {journal} {\apj}\ }\textbf {\bibinfo {volume} {584}},\ \bibinfo
  {pages} {971} (\bibinfo {year} {2003})},\ \Eprint
  {http://arxiv.org/abs/astro-ph/0210634} {astro-ph/0210634} \BibitemShut
  {NoStop}%
\bibitem [{\citenamefont {{Foglizzo}}\ \emph {et~al.}(2007)\citenamefont
  {{Foglizzo}}, \citenamefont {{Galletti}}, \citenamefont {{Scheck}},\ and\
  \citenamefont {{Janka}}}]{foglizzo:07}%
  \BibitemOpen
  \bibfield  {author} {\bibinfo {author} {\bibfnamefont {T.}~\bibnamefont
  {{Foglizzo}}}, \bibinfo {author} {\bibfnamefont {P.}~\bibnamefont
  {{Galletti}}}, \bibinfo {author} {\bibfnamefont {L.}~\bibnamefont
  {{Scheck}}}, \ and\ \bibinfo {author} {\bibfnamefont {H.-T.}\ \bibnamefont
  {{Janka}}},\ }\href {\doibase 10.1086/509612} {\bibfield  {journal} {\bibinfo
   {journal} {\apj}\ }\textbf {\bibinfo {volume} {654}},\ \bibinfo {pages}
  {1006} (\bibinfo {year} {2007})},\ \Eprint
  {http://arxiv.org/abs/astro-ph/0606640} {astro-ph/0606640} \BibitemShut
  {NoStop}%
\bibitem [{\citenamefont {{Burrows}}\ \emph {et~al.}(2006)\citenamefont
  {{Burrows}}, \citenamefont {{Reddy}},\ and\ \citenamefont
  {{Thompson}}}]{burrows:06}%
  \BibitemOpen
  \bibfield  {author} {\bibinfo {author} {\bibfnamefont {A.}~\bibnamefont
  {{Burrows}}}, \bibinfo {author} {\bibfnamefont {S.}~\bibnamefont {{Reddy}}},
  \ and\ \bibinfo {author} {\bibfnamefont {T.~A.}\ \bibnamefont {{Thompson}}},\
  }\href {\doibase 10.1016/j.nuclphysa.2004.06.012} {\bibfield  {journal}
  {\bibinfo  {journal} {Nuclear Physics A}\ }\textbf {\bibinfo {volume}
  {777}},\ \bibinfo {pages} {356} (\bibinfo {year} {2006})},\ \Eprint
  {http://arxiv.org/abs/astro-ph/0404432} {astro-ph/0404432} \BibitemShut
  {NoStop}%
\bibitem [{\citenamefont {{Bruenn}}(1985)}]{1985ApJS...58..771B}%
  \BibitemOpen
  \bibfield  {author} {\bibinfo {author} {\bibfnamefont {S.~W.}\ \bibnamefont
  {{Bruenn}}},\ }\href {\doibase 10.1086/191056} {\bibfield  {journal}
  {\bibinfo  {journal} {\apjs}\ }\textbf {\bibinfo {volume} {58}},\ \bibinfo
  {pages} {771} (\bibinfo {year} {1985})}\BibitemShut {NoStop}%
\bibitem [{\citenamefont {{Thompson}}\ \emph {et~al.}(2000)\citenamefont
  {{Thompson}}, \citenamefont {{Burrows}},\ and\ \citenamefont
  {{Horvath}}}]{thomp_bur_horvath}%
  \BibitemOpen
  \bibfield  {author} {\bibinfo {author} {\bibfnamefont {T.~A.}\ \bibnamefont
  {{Thompson}}}, \bibinfo {author} {\bibfnamefont {A.}~\bibnamefont
  {{Burrows}}}, \ and\ \bibinfo {author} {\bibfnamefont {J.~E.}\ \bibnamefont
  {{Horvath}}},\ }\href {\doibase 10.1103/PhysRevC.62.035802} {\bibfield
  {journal} {\bibinfo  {journal} {\prc}\ }\textbf {\bibinfo {volume} {62}},\
  \bibinfo {eid} {035802} (\bibinfo {year} {2000})},\ \Eprint
  {http://arxiv.org/abs/astro-ph/0003054} {astro-ph/0003054} \BibitemShut
  {NoStop}%
\bibitem [{\citenamefont {{Rampp}}\ and\ \citenamefont
  {{Janka}}(2002)}]{rampp_janka2002}%
  \BibitemOpen
  \bibfield  {author} {\bibinfo {author} {\bibfnamefont {M.}~\bibnamefont
  {{Rampp}}}\ and\ \bibinfo {author} {\bibfnamefont {H.-T.}\ \bibnamefont
  {{Janka}}},\ }\href {\doibase 10.1051/0004-6361:20021398} {\bibfield
  {journal} {\bibinfo  {journal} {\aap}\ }\textbf {\bibinfo {volume} {396}},\
  \bibinfo {pages} {361} (\bibinfo {year} {2002})},\ \Eprint
  {http://arxiv.org/abs/astro-ph/0203101} {astro-ph/0203101} \BibitemShut
  {NoStop}%
\bibitem [{\citenamefont {{Burrows}}\ and\ \citenamefont
  {{Thompson}}(2002)}]{2002astro.ph.11404B}%
  \BibitemOpen
  \bibfield  {author} {\bibinfo {author} {\bibfnamefont {A.}~\bibnamefont
  {{Burrows}}}\ and\ \bibinfo {author} {\bibfnamefont {T.~A.}\ \bibnamefont
  {{Thompson}}},\ }\href@noop {} {\bibfield  {journal} {\bibinfo  {journal}
  {ArXiv Astrophysics e-prints}\ } (\bibinfo {year} {2002})},\ \Eprint
  {http://arxiv.org/abs/astro-ph/0211404} {astro-ph/0211404} \BibitemShut
  {NoStop}%
\bibitem [{\citenamefont {{Burrows}}\ and\ \citenamefont
  {{Thompson}}(2004)}]{burrows_thompson2004}%
  \BibitemOpen
  \bibfield  {author} {\bibinfo {author} {\bibfnamefont {A.}~\bibnamefont
  {{Burrows}}}\ and\ \bibinfo {author} {\bibfnamefont {T.~A.}\ \bibnamefont
  {{Thompson}}},\ }in\ \href {\doibase 10.1007/978-0-306-48599-2_5} {\emph
  {\bibinfo {booktitle} {Astrophysics and Space Science Library}}},\ \bibinfo
  {series} {Astrophysics and Space Science Library}, Vol.\ \bibinfo {volume}
  {302},\ \bibinfo {editor} {edited by\ \bibinfo {editor} {\bibfnamefont
  {C.~L.}\ \bibnamefont {{Fryer}}}}\ (\bibinfo {year} {2004})\ pp.\ \bibinfo
  {pages} {133--174}\BibitemShut {NoStop}%
\bibitem [{\citenamefont {{Bruenn}}\ \emph {et~al.}(2018)\citenamefont
  {{Bruenn}}, \citenamefont {{Blondin}}, \citenamefont {{Hix}}, \citenamefont
  {{Lentz}}, \citenamefont {{Messer}}, \citenamefont {{Mezzacappa}},
  \citenamefont {{Endeve}}, \citenamefont {{Harris}}, \citenamefont
  {{Marronetti}}, \citenamefont {{Budiardja}}, \citenamefont {{Chertkow}},\
  and\ \citenamefont {{Lee}}}]{chimera}%
  \BibitemOpen
  \bibfield  {author} {\bibinfo {author} {\bibfnamefont {S.~W.}\ \bibnamefont
  {{Bruenn}}}, \bibinfo {author} {\bibfnamefont {J.~M.}\ \bibnamefont
  {{Blondin}}}, \bibinfo {author} {\bibfnamefont {W.~R.}\ \bibnamefont
  {{Hix}}}, \bibinfo {author} {\bibfnamefont {E.~J.}\ \bibnamefont {{Lentz}}},
  \bibinfo {author} {\bibfnamefont {O.~E.~B.}\ \bibnamefont {{Messer}}},
  \bibinfo {author} {\bibfnamefont {A.}~\bibnamefont {{Mezzacappa}}}, \bibinfo
  {author} {\bibfnamefont {E.}~\bibnamefont {{Endeve}}}, \bibinfo {author}
  {\bibfnamefont {J.~A.}\ \bibnamefont {{Harris}}}, \bibinfo {author}
  {\bibfnamefont {P.}~\bibnamefont {{Marronetti}}}, \bibinfo {author}
  {\bibfnamefont {R.~D.}\ \bibnamefont {{Budiardja}}}, \bibinfo {author}
  {\bibfnamefont {M.~A.}\ \bibnamefont {{Chertkow}}}, \ and\ \bibinfo {author}
  {\bibfnamefont {C.-T.}\ \bibnamefont {{Lee}}},\ }\href@noop {} {\bibfield
  {journal} {\bibinfo  {journal} {arXiv e-prints}\ ,\ \bibinfo {eid}
  {arXiv:1809.05608}} (\bibinfo {year} {2018})},\ \Eprint
  {http://arxiv.org/abs/1809.05608} {arXiv:1809.05608 [astro-ph.IM]}
  \BibitemShut {NoStop}%
\bibitem [{\citenamefont {{Vartanyan}}\ \emph {et~al.}()\citenamefont
  {{Vartanyan}}, \citenamefont {{Burrows}}, \citenamefont {{Radice}},
  \citenamefont {{Skinner}},\ and\ \citenamefont {{Dolence}}}]{vartanyan2018a}%
  \BibitemOpen
  \bibfield  {author} {\bibinfo {author} {\bibfnamefont {D.}~\bibnamefont
  {{Vartanyan}}}, \bibinfo {author} {\bibfnamefont {A.}~\bibnamefont
  {{Burrows}}}, \bibinfo {author} {\bibfnamefont {D.}~\bibnamefont {{Radice}}},
  \bibinfo {author} {\bibfnamefont {M.~A.}\ \bibnamefont {{Skinner}}}, \ and\
  \bibinfo {author} {\bibfnamefont {J.}~\bibnamefont {{Dolence}}},\ }\href@noop
  {} {\ }\BibitemShut {NoStop}%
\bibitem [{\citenamefont {{Suwa}}\ \emph {et~al.}(2019)\citenamefont {{Suwa}},
  \citenamefont {{Tahara}},\ and\ \citenamefont {{Komatsu}}}]{Suwa2019}%
  \BibitemOpen
  \bibfield  {author} {\bibinfo {author} {\bibfnamefont {Y.}~\bibnamefont
  {{Suwa}}}, \bibinfo {author} {\bibfnamefont {H.~W.~H.}\ \bibnamefont
  {{Tahara}}}, \ and\ \bibinfo {author} {\bibfnamefont {E.}~\bibnamefont
  {{Komatsu}}},\ }\href {\doibase 10.1093/ptep/ptz087} {\bibfield  {journal}
  {\bibinfo  {journal} {Progress of Theoretical and Experimental Physics}\
  }\textbf {\bibinfo {volume} {2019}},\ \bibinfo {eid} {083E04} (\bibinfo
  {year} {2019})},\ \Eprint {http://arxiv.org/abs/1904.05047} {arXiv:1904.05047
  [astro-ph.HE]} \BibitemShut {NoStop}%
\bibitem [{\citenamefont {{Kompaneets}}(1957)}]{kom1957}%
  \BibitemOpen
  \bibfield  {author} {\bibinfo {author} {\bibfnamefont {A.~S.}\ \bibnamefont
  {{Kompaneets}}},\ }\href@noop {} {\bibfield  {journal} {\bibinfo  {journal}
  {Soviet Journal of Experimental and Theoretical Physics}\ }\textbf {\bibinfo
  {volume} {4}},\ \bibinfo {pages} {730} (\bibinfo {year} {1957})}\BibitemShut
  {NoStop}%
\bibitem [{\citenamefont {{Sunyaev}}\ and\ \citenamefont
  {{Zeldovich}}(1972)}]{sunyaev1972}%
  \BibitemOpen
  \bibfield  {author} {\bibinfo {author} {\bibfnamefont {R.~A.}\ \bibnamefont
  {{Sunyaev}}}\ and\ \bibinfo {author} {\bibfnamefont {Y.~B.}\ \bibnamefont
  {{Zeldovich}}},\ }\href@noop {} {\bibfield  {journal} {\bibinfo  {journal}
  {Comments on Astrophysics and Space Physics}\ }\textbf {\bibinfo {volume}
  {4}},\ \bibinfo {pages} {173} (\bibinfo {year} {1972})}\BibitemShut {NoStop}%
\bibitem [{\citenamefont {{Rybicki}}\ and\ \citenamefont
  {{Lightman}}(1979)}]{rybicki_1979}%
  \BibitemOpen
  \bibfield  {author} {\bibinfo {author} {\bibfnamefont {G.~B.}\ \bibnamefont
  {{Rybicki}}}\ and\ \bibinfo {author} {\bibfnamefont {A.~P.}\ \bibnamefont
  {{Lightman}}},\ }\href@noop {} {\emph {\bibinfo {title} {{Radiative processes
  in astrophysics}}}}\ (\bibinfo {year} {1979})\BibitemShut {NoStop}%
\bibitem [{\citenamefont {Olive}(2016)}]{PDG_2016}%
  \BibitemOpen
  \bibfield  {author} {\bibinfo {author} {\bibfnamefont {K.}~\bibnamefont
  {Olive}},\ }\href {\doibase 10.1088/1674-1137/40/10/100001} {\bibfield
  {journal} {\bibinfo  {journal} {Chinese Physics C}\ }\textbf {\bibinfo
  {volume} {40}},\ \bibinfo {pages} {100001} (\bibinfo {year}
  {2016})}\BibitemShut {NoStop}%
\bibitem [{\citenamefont {{Horowitz}}(2002)}]{horowitz2002}%
  \BibitemOpen
  \bibfield  {author} {\bibinfo {author} {\bibfnamefont {C.~J.}\ \bibnamefont
  {{Horowitz}}},\ }\href {\doibase 10.1103/PhysRevD.65.043001} {\bibfield
  {journal} {\bibinfo  {journal} {\prd}\ }\textbf {\bibinfo {volume} {65}},\
  \bibinfo {eid} {043001} (\bibinfo {year} {2002})},\ \Eprint
  {http://arxiv.org/abs/astro-ph/0109209} {astro-ph/0109209} \BibitemShut
  {NoStop}%
\bibitem [{\citenamefont {{Melson}}\ \emph
  {et~al.}(2015{\natexlab{b}})\citenamefont {{Melson}}, \citenamefont
  {{Janka}}, \citenamefont {{Bollig}}, \citenamefont {{Hanke}}, \citenamefont
  {{Marek}},\ and\ \citenamefont {{M{\"u}ller}}}]{melson:15b}%
  \BibitemOpen
  \bibfield  {author} {\bibinfo {author} {\bibfnamefont {T.}~\bibnamefont
  {{Melson}}}, \bibinfo {author} {\bibfnamefont {H.-T.}\ \bibnamefont
  {{Janka}}}, \bibinfo {author} {\bibfnamefont {R.}~\bibnamefont {{Bollig}}},
  \bibinfo {author} {\bibfnamefont {F.}~\bibnamefont {{Hanke}}}, \bibinfo
  {author} {\bibfnamefont {A.}~\bibnamefont {{Marek}}}, \ and\ \bibinfo
  {author} {\bibfnamefont {B.}~\bibnamefont {{M{\"u}ller}}},\ }\href {\doibase
  10.1088/2041-8205/808/2/L42} {\bibfield  {journal} {\bibinfo  {journal}
  {\apjl}\ }\textbf {\bibinfo {volume} {808}},\ \bibinfo {eid} {L42} (\bibinfo
  {year} {2015}{\natexlab{b}})},\ \Eprint {http://arxiv.org/abs/1504.07631}
  {arXiv:1504.07631 [astro-ph.SR]} \BibitemShut {NoStop}%
\bibitem [{\citenamefont {Ahmed}(2012)}]{ahmed2012}%
  \BibitemOpen
  \bibfield  {author} {\bibinfo {author} {\bibfnamefont {A.~K. A. A. D. S.
  e.~a.}\ \bibnamefont {Ahmed}, \bibfnamefont {Z.}} (\bibinfo {collaboration}
  {HAPPEX Collaboration}),\ }\href@noop {} {\bibfield  {journal} {\bibinfo
  {journal} {Phys. Rev. Lett.}\ }\textbf {\bibinfo {volume} {108}},\ \bibinfo
  {pages} {102001} (\bibinfo {year} {2012})}\BibitemShut {NoStop}%
\bibitem [{\citenamefont {{Green}}\ \emph {et~al.}(2017)\citenamefont
  {{Green}}, \citenamefont {{Hasan}}, \citenamefont {{Meinel}}, \citenamefont
  {{Engelhardt}}, \citenamefont {{Krieg}}, \citenamefont {{Laeuchli}},
  \citenamefont {{Negele}}, \citenamefont {{Orginos}}, \citenamefont
  {{Pochinsky}},\ and\ \citenamefont {{Syritsyn}}}]{green2017}%
  \BibitemOpen
  \bibfield  {author} {\bibinfo {author} {\bibfnamefont {J.}~\bibnamefont
  {{Green}}}, \bibinfo {author} {\bibfnamefont {N.}~\bibnamefont {{Hasan}}},
  \bibinfo {author} {\bibfnamefont {S.}~\bibnamefont {{Meinel}}}, \bibinfo
  {author} {\bibfnamefont {M.}~\bibnamefont {{Engelhardt}}}, \bibinfo {author}
  {\bibfnamefont {S.}~\bibnamefont {{Krieg}}}, \bibinfo {author} {\bibfnamefont
  {J.}~\bibnamefont {{Laeuchli}}}, \bibinfo {author} {\bibfnamefont
  {J.}~\bibnamefont {{Negele}}}, \bibinfo {author} {\bibfnamefont
  {K.}~\bibnamefont {{Orginos}}}, \bibinfo {author} {\bibfnamefont
  {A.}~\bibnamefont {{Pochinsky}}}, \ and\ \bibinfo {author} {\bibfnamefont
  {S.}~\bibnamefont {{Syritsyn}}},\ }\href {\doibase
  10.1103/PhysRevD.95.114502} {\bibfield  {journal} {\bibinfo  {journal}
  {\prd}\ }\textbf {\bibinfo {volume} {95}},\ \bibinfo {eid} {114502} (\bibinfo
  {year} {2017})},\ \Eprint {http://arxiv.org/abs/1703.06703} {arXiv:1703.06703
  [hep-lat]} \BibitemShut {NoStop}%
\end{thebibliography}%


%

\clearpage

\appendix
\section{Properties of $g^{(k)}(x,\theta,\theta')$}
\label{app:A}
First, let's consider the isotropic case.
Recall that $g(x;x')=(1-f)f'e^{-\frac{1}{2}(x-x')}-f(1-f')e^{\frac{1}{2}(x-x')}$ and $g^{(k)}(x)=\frac{\partial^k g(x;x')}{\partial x'^k}\vert_{x'=x}$. The first few $g^{(k)}(x)$ are:
\equ{
g^{(0)}(x)&=&0\\
g^{(1)}(x)&=&\frac{df}{dx}+f-f^2\\
g^{(2)}(x)&=&\frac{d^2f}{dx^2}+\frac{df}{dx}-2f\frac{df}{dx}\\
g^{(3)}(x)&=&\frac{d^3f}{dx^3}+\frac{3}{2}(1-2f)\frac{d^2f}{dx^2}+\frac{3}{4}\frac{df}{dx}\nonumber\\
& &+\frac{1}{4}f(1-f)\\
g^{(4)}(x)&=&\frac{d^4f}{dx^4}+2(1-2f)\frac{d^3f}{dx^3}+\frac{3}{2}\frac{d^2f}{dx^2}\nonumber\\
& &+\frac{1}{2}(1-2f)\frac{df}{dx}
\label{equ:gk}
}

In the anisotropic case, $g^{(k)}=\frac{\partial^k g(x,\theta;x',\theta')}{\partial x'^k}\vert_{x'=x}$ are more complex:
\equ{
g^{(0)}(x,\theta,\theta')&=&f(x,\theta')-f(x,\theta)\\
g^{(1)}(x,\theta,\theta')&=&\frac{df(x,\theta')}{dx}+\frac{1}{2}\left[f(x,\theta)+f(x,\theta')\right]\nonumber\\
& &-f(x,\theta)f(x,\theta')\\
g^{(2)}(x,\theta,\theta')&=&\frac{d^2f(x,\theta')}{dx^2}+\frac{df(x,\theta')}{dx}\nonumber\\
& &-2f(x,\theta)\frac{df(x,\theta')}{dx}\nonumber\\
& &+\frac{1}{4}\left[f(x,\theta')-f(x,\theta)\right]\\
g^{(3)}(x,\theta,\theta')&=&\frac{d^3f(x,\theta')}{dx^3}+\frac{3}{2}\left[1-2f(x,\theta)\right]\frac{d^2f(x,\theta')}{dx^2}\nonumber\\
& &+\frac{3}{4}\frac{df(x,\theta')}{dx}+\frac{1}{8}\left[f(x,\theta')+f(x,\theta)\right]\nonumber\\
& &-\frac{1}{4}f(x,\theta')f(x,\theta)\\
g^{(4)}(x,\theta,\theta')&=&\frac{d^4f(x,\theta')}{dx^4}+2\left[1-2f(x,\theta)\right]\frac{d^3f(x,\theta')}{dx^3}\nonumber\\
& &+\frac{3}{2}\frac{d^2f(x,\theta')}{dx^2}+\frac{1}{2}\left[1-2f(x,\theta)\right]\frac{df(x,\theta')}{dx}\nonumber\\
& &+\frac{1}{16}\left[f(x,\theta')-f(x,\theta)\right]\, .
}
However, if we integrate $\theta$ and $\theta'$ out symmetrically, all antisymmetric terms such as $f(x,\theta')-f(x,\theta)$ vanish and symmetric terms such as $f(x,\theta)+f(x,\theta')$ can be replaced by either $2f(x,\theta)$ or $2f(x,\theta')$. This property can be used to simplify the calculation.

\section{Calculation of $a_k(x,\mu)$}\label{sec:ak}
The definition of $a_k(x,\mu)$ is
\wide{
\equ{
a_k(x,\mu)&=&\int_{-\infty}^1\frac{(-x\alpha)^k}{k!}\tilde{S}(x,x(1-\alpha),\mu)(1-\alpha)^2d\alpha\\
\tilde{S}(x,x(1-\alpha),\mu)
&=&\frac{n_N(2\pi\beta m)^{\frac{1}{2}}\beta}{x\sqrt{1+(1-\alpha)^2-2(1-\alpha)\mu}}\nonumber\\
& &\times\exp\{-\frac{\beta m}{2}[\frac{\alpha^2}{1+(1-\alpha)^2-2(1-\alpha)\mu}+\frac{1+(1-\alpha)^2-2(1-\alpha)\mu}{4\beta^2m^2}x^2]\}\, .
}
}
Since we only care here about non-relativistic nucleons, we know that $\beta m=\frac{m}{kT}\gg1$ and, thus, that $\tilde{S}(x,x(1-\alpha),\mu)$ decays very fast as $\alpha$ increases. When $\alpha\geq1$, the exponential term in $\tilde{S}$ is smaller than $\exp(-\frac{\beta m}{2})\ll1$. Therefore, we can extend the upper bound of the integral to $+\infty$:
\equ{
a_k(x,\mu)&=&\int_{-\infty}^\infty d\alpha\frac{(-x\alpha)^k}{k!}(1-\alpha)^2\nonumber\\
& &\times \tilde{S}(x,x(1-\alpha),\mu)\, .
}
We calculate these coefficients using Gaussian integrals. We first split the Gaussian function term from the exponential term and then expand the remaining part as polynomials. Then, we integrate out these Gaussian integrals. Since the Gaussian part is $\exp\{-\frac{\beta m}{2}\frac{\alpha^2}{2-2\mu}\}$, each $\alpha$ in front of it will be of order $\frac{1}{\sqrt{\beta m}}$ after the integration. Therefore, we can truncate at some order of $\alpha$. However, the derivative of $\tilde{S}$ can sometimes introduce extra $\beta m$s, so one should be very careful when throwing away higher-order terms.

The function to be expanded is
\wide{
\equ{
\mathcal{S}&=&\tilde{S}(x,x(1-\alpha),\mu)(1-\alpha)^2\exp\{\frac{\beta m}{2}\frac{\alpha^2}{2-2\mu}\}\nonumber\\
&=&\frac{C(1-\alpha)^2}{\sqrt{1+(1-\alpha)^2-2(1-\alpha)\mu}}\nonumber\\
& &\times\exp\{-\frac{\beta m}{2}[\frac{\alpha^2}{1+(1-\alpha)^2-2(1-\alpha)\mu}+\frac{1+(1-\alpha)^2-2(1-\alpha)\mu}{4\beta^2m^2}x^2-\frac{\alpha^2}{2-2\mu}]\}\, ,
}
}
where $C=\frac{n_N(2\pi\beta m)^{\frac{1}{2}}\beta}{x}$ is a constant. We expand this function around $\alpha=0$ and put terms that are the same order in $\frac{1}{\beta m}$ together. This can be done by a small trick: let $\alpha=\frac{t}{\sqrt{\beta m}}$ and expand $\mathcal{S}$ in terms of $\frac{1}{\sqrt{\beta m}}$. After truncating at some order of $\frac{1}{\sqrt{\beta m}}$, we replace $t$ with $\alpha\sqrt{\beta m}$.
\wide{
\equ{
\mathcal{S}&=&C\{\frac{1}{\sqrt{2}(1-\mu)^{1/2}}+\frac{-6(1-\mu)\alpha-(\beta m)\alpha^3}{4\sqrt{2}(1-\mu)^{3/2}}+\frac{(\beta m)^3\alpha^6+4(\beta m)^2\alpha^4(2-\mu)+4(\beta m)\alpha^2(1-4\mu+3\mu^2)-8x^2(1-\mu)^3}{32\sqrt{2}\beta m(1-\mu)^{5/2}}\nonumber\\
& &-\frac{(\beta m)^4\alpha^9+6(\beta m)^3(1+\mu)\alpha^7-12(\beta m)^2(5-6\mu+\mu^2)\alpha^5-24(\beta m)(1-\mu)^2(3-\mu+x^2(1-\mu))\alpha^3-240x^2(1-\mu)^4\alpha}{384\sqrt{2}(1-\mu)^{7/2}}\nonumber\\
& &+O((\frac{1}{\sqrt{\beta m}})^4)\}\, .
}
}
Then, we can calculate the coefficients:
\wide{
\equ{
a_0(x,\mu)&\approx&\frac{n_N\pi}{mx}[\frac{4\beta m- x^2 (1- \mu)  +28  - 24 \mu}{2}+O(\frac{1}{\beta m})]\label{equ:ak0}\\
a_1(x,\mu)&\approx&\frac{n_N\pi}{mx}[\frac{x(12\beta m- 4x^2 (1 - \mu) +24(7-4\mu)) (1- \mu) }{\beta m}+O((\frac{1}{\beta m})^2)]\label{equ:ak1}\\
a_2(x,\mu)&\approx&\frac{n_N\pi}{mx}[\frac{x^2 (4 \beta m + 24 (7 - 6 \mu) -
    x^2 (1-\mu)) (1-\mu) }{2 \beta m}+O((\frac{1}{\beta m})^2)]\label{equ:ak2}\\
a_3(x,\mu)&\approx&\frac{n_N\pi}{mx}[\frac{ 16 x^3 (1-\mu)^2}{\beta m}+O((\frac{1}{\beta m})^2)]\label{equ:ak3}\\
a_4(x,\mu)&\approx&\frac{n_N\pi}{mx}[\frac{ x^4 (1-\mu)^2}{\beta m}+O((\frac{1}{\beta m})^2)]\, .
\label{equ:ak4}
}
}
Although $a_0$ has a larger residual than other coefficients, the vanishing $g^{(0)}(x;x)$ will erase such residuals.

\end{document}